\documentclass{mn2e}
\usepackage{epsfig}
\usepackage{epsf}

\title[Core-Collapse Supernova Fractions]{Observed Fractions of
  Core-Collapse Supernova Types and Initial Masses of their Single and
  Binary Progenitor Stars}

\author[N.\ Smith et al.]{Nathan Smith$^1$\thanks{Email:
    nathans@astro.berkeley.edu}, Weidong Li$^{1}$, Alexei V.\
  Filippenko$^{1}$, \& Ryan Chornock$^2$ \\ $^1$ Department of Astronomy,
  University of California, Berkeley, CA 94720-3411, USA \\ $^2$
  Harvard-Smithsonian Center for Astrophysics, 60 Garden St.,
  Cambridge, MA 02138, USA}

\begin{document}
\date{Accepted 0000, Received 0000, in original form 0000}
\pagerange{\pageref{firstpage}--\pageref{lastpage}} \pubyear{2009}
\def\arcdeg{\degr}
\maketitle
\label{firstpage}

\begin{abstract}

  We analyse the observed fractions of core-collapse supernova (SN)
  types from the Lick Observatory SN Search (LOSS), and we discuss the
  corresponding implications for massive star evolution.  For a
  standard initial mass function, observed fractions of SN types
  cannot be reconciled with the expectations of single-star evolution.
  The mass range of Wolf-Rayet (WR) stars that shed their hydrogen
  envelopes via their own mass loss accounts for less than half of the
  observed fraction of Type~Ibc supernovae (SNe~Ibc).  The true
  progenitors of SNe~Ibc must extend to a much lower range of initial
  masses than classical WR stars, and we argue that most SN~Ibc and
  SN~IIb progenitors must arise from binary Roche-lobe overflow.  In
  this scenario, SNe~Ic would still trace higher initial mass and
  metallicity, because line-driven winds in the WR stage remove the
  helium layer and propel the transition from SN~Ib to Ic.  Less
  massive progenitors of SNe~Ib and IIb may not be classical WR stars;
  they may be underluminous with weak winds, possibly hidden by
  overluminous mass-gainer companions that could appear as B[e]
  supergiants or related objects having aspherical circumstellar
  material.  The remaining SN types (II-P, II-L, and IIn) need to be
  redistributed across the full range of initial masses, so that even
  some very massive single stars retain H envelopes until explosion.
  We consider the possibility of direct collapse to black holes
  without visible SNe, but find this hypothesis difficult to
  accommodate in most scenarios.  Major areas of remaining uncertainty
  are (1) the detailed influence of binary separation, rotation, and
  metallicity, (2) mass differences in progenitors of SNe~IIn compared
  to SNe~II-L and II-P, and (3) the fraction of SNe~Ic arising from
  single stars with the help of eruptive mass loss, how this depends
  on metallicity, and how it relates to diversity within the SN~Ic
  subclass.  Continued studies of progenitor stars and their
  environments in nearby galaxies, accounting for SN types, may
  eventually test these ideas.

\end{abstract}


\section{INTRODUCTION}

The observed fractions of various types of core-collapse supernovae
(CCSNe) provide key information about the evolution and ultimate fates
of massive stars.  Because of their tremendous luminosity, SNe can
potentially be used as diagnostics of mass loss and the evolution of
individual stars at great distances and in a variety of galactic
environments, but only if we first understand how to map initial
masses and evolution of different progenitor stars to the various
types of SNe that they produce.  If drawn from a stellar population
that obeys a standard initial mass function (IMF), the observed
fractions of different CCSN types constrain the ranges of initial mass
for their progenitors, as well as the evolutionary paths they take
before death.  The aim of this paper is to explore how the IMF can be
sampled in order to be consistent with the observed fractions of CCSN
types.


The main observed types of SNe that we consider are II-P (plateau),
II-L (linear), IIn (relatively narrow lines), IIb (transitional), Ib,
Ic, and Ibc-pec (see \S 2).  Spectroscopic classification criteria for
these are reviewed by Filippenko (1997).  Pre-SN mass loss of the
progenitor star determines which of these types of SN is seen,
stripping away various amounts of the star's H and possibly He
envelopes before core collapse ejects the remaining envelope.  SNe~Ibc
are the result of complete removal of the H envelope, SNe~IIb have
retained only a small H mass (typically $<$0.5 M$_{\odot}$), while
SNe~II-L and II-P have retained increasingly more of their H
envelopes.  SNe~IIn are different in the sense that their spectral
appearance is determined largely by shock interaction with
circumstellar material (CSM) lost in the decades or centuries
preceding core collapse.

The three potential mechanisms for a SN progenitor's mass loss are via
steady winds, eruptive mass loss, or mass transfer due to Roche-lobe
overflow (RLOF) in a close binary system.  Depending on which
dominates, the amount of mass lost may or may not depend in a simple
way on metallicity or on the initial mass of the progenitor star,
making reliable predictions difficult without a more complete
understanding of mass loss.  The evolutionary state --- red supergiant
(RSG), blue supergiant (BSG), luminous blue variable (LBV), and
Wolf-Rayet (WR) stars of the WN and WC sequences --- and hence the
stellar radius at the time of explosion, are also important, although
these can be considered as largely the {\it result} of mass loss.
Massive stars have substantial steady stellar winds through most of
their lives (see Lamers \& Cassinelli 1999), with either
metallicity-dependent, line-driven winds in hot stars (Kudritzki \&
Puls 2000), or slow, pulsation/dust-driven winds in cool stars
(Reimers 1977).  Single-star evolution models adopt simple
prescriptions for these steady winds (Meynet et al.\ 1994), and aim to
predict the fates of massive stars as functions of initial mass and
metallicity (e.g., Heger et al.\ 2003; Eldridge \& Tout 2004).

Recent observational work, however, has demonstrated that the standard
observational mass-loss rates used as input to these models are far
too high; the standard mass-loss rates of hot stars (e.g.,
Nieuwenhuijzen \& de Jager 1990; de Jager et al.\ 1998) are reduced by
factors of 3--10 when the effects of clumping are considered properly
(Bouret et al.\ 2005; Fullerton et al.\ 2006; Puls et al.\ 2006).
There is also a parallel problem in cool star mass-loss
rates --- reduction of an order of magnitude to the standard Reimers
formula for red giants may be required (M\'{e}sz\'{a}ros et al.\
2009), and it would be interesting if this also affects more massive
RSG stars. These lower mass-loss rates have a profound impact on stellar
evolution and SN progenitors, requiring us to turn to either eruptive
mass loss (Smith \& Owocki 2006) or close binaries (e.g.,
Paczy\'{n}ski 1967) to make up the deficit.  We will see that this
turns out to be a major theme in explaining the frequencies of SN
types.


The stripping of a star's H envelope due to mass transfer in RLOF
binary systems has long been considered a likely mechanism to produce
WR stars and the progenitors of SNe~Ibc (e.g., Paczy\'{n}ski 1967;
Podsiadlowski et al.\ 1992).  Recent stellar evolution models attempt
to account for this (e.g., Eldridge et al.\ 2008), but considerable
uncertainty surrounds empirical estimates of binary fractions; see
Kobulnicky \& Fryer (2007) and references therein.

In addition to close binary evolution, a major uncertainty concerns
the net effect of episodic and eruptive mass loss during late stages
of stellar evolution (Smith \& Owocki 2006).  These outbursts are
observed to occur, and studies suggest that they shed more mass
from a star than do steady winds (Smith \& Owocki 2006).  The importance
of sudden, short-duration eruptive mass loss is a concern for the
predictive power of any stellar evolution model, none of which
currently include it.  Observational clues from CSM interaction in
SNe~IIn dictate that heavy mass loss sometimes occurs shortly before
core collapse (e.g., Smith et al.\ 2007, 2008b, 2010; Chugai et al.\
2004); if heavy mass loss is concentrated in brief events during the
last few thousand years before core collapse, then the statistical
distribution of end fates (i.e., SN types) won't necessarily reflect
the observed relative fractions of WN, WC, RSG, LBVs, and so on, which are
determined by the time spent in each state.  This is critical and
potentially misleading, since many stellar evolution codes are linked
to these observed fractions.

Another key point is that both binaries and eruptions are probably
less sensitive to metallicity than line-driven winds of hot stars.
Some studies have shown that the observed fraction of SNe~Ibc compared
to SNe~II increases with metallicity, implying that 
metallicity-dependent winds play an important role (Prantzos \& Boissier 
2003; Prieto et al.\ 2008; Boissier \& Prantzos 2009).  On the other 
hand, observations have also revealed a large population of WR stars 
in low-metallicity galaxies, which cannot be explained by stellar 
winds alone (Izotov et al.\ 1997; Brown et al.\ 2002; Crowther \& 
Hadfield 2006), while the broad-lined SNe~Ic that accompany gamma-ray 
bursts (see Woosley \& Bloom 2006 for a review) seem to prefer low 
metallicity (Stanek et al.\ 2006; Modjaz et al.\ 2008).


There have been a few previous investigations of relative SN rates
that our study builds upon.  Cappellaro et al.\ (1997) examined the
statistics from 110 SNe (including SNe~Ia), deriving widely adopted
rates of various SN types in different environments.  More recently,
Smartt et al.\ (2009) considered a volume-limited sample of nearby
CCSNe and examined the relative fractions of SNe II-P, II-L, IIb, IIn,
Ib, and Ic, as we do.  Our results are different from theirs, as
described below, leading to some quite different implications for
massive stars.  Finally, Arcavi et al.\ (2010) have recently submitted
a paper independent of our study using SNe from the Palomar Transient
Factory (PTF), finding a difference in the relative fractions of SNe
II, IIb, Ib, Ic, and broad-lined Ic between large galaxies and dwarf
galaxies.  Since our survey did not adequately sample dwarf galaxies,
the study by Arcavi et al.\ (2010) is complementary to ours, although
we find significantly different relative fractions of SN types in
large galaxies.  We also consider direct detections of SN progenitor
stars from pre-explosion data, discussed in considerable detail below
and reviewed recently by Smartt (2009).  Throughout, we include this
information along with current ideas about massive single and binary
stars.


Here we present and discuss the implications of the observed relative
fractions of different types of CCSNe in a new volume-limited sample,
measured during the course of the Lick Observatory Supernova Search
(LOSS) conducted with the Katzman Automatic Imaging Telescope (KAIT;
Filippenko et al. 2001).  This follows a series of papers discussing
LOSS.  Paper I (Leaman et al.\ 2010) describes the method of deriving
rates from LOSS data, Paper II (Li et al.\ 2010a) discusses the
luminosity functions of SNe and gives a detailed discussion of how the
different fractions of SN types were derived, and Paper III (Li et
al.\ 2010b) presents relations with host galaxies and other details.

\begin{table}\begin{minipage}{3.1in}\begin{center}
\caption{Volume-limited core-collapse SN fractions}\scriptsize
\begin{tabular}{@{}lrc}\hline\hline
SN Type  &fraction &error \\ \hline
         &( \% )   &( \% ) \\ \hline
Ic       &14.9   &+4.2/$-$3.8 \\
Ib       &7.1    &+3.1/$-$2.6 \\
Ibc-pec  &4.0    &+2.0/$-$2.4 \\
IIb      &10.6   &+3.6/$-$3.1 \\
IIn      &8.8    &+3.3/$-$2.9 \\
II-L     &6.4    &+2.9/$-$2.5 \\
II-P     &48.2   &+5.7/$-$5.6 \\ \hline
Ibc (all)&26.0   &+5.1/$-$4.8 \\
Ibc+IIb  &36.5   &+5.5/$-$5.4 \\
\hline
\end{tabular}
\end{center}\end{minipage}
\end{table}

\begin{figure}\begin{center}
\includegraphics[width=3.0in]{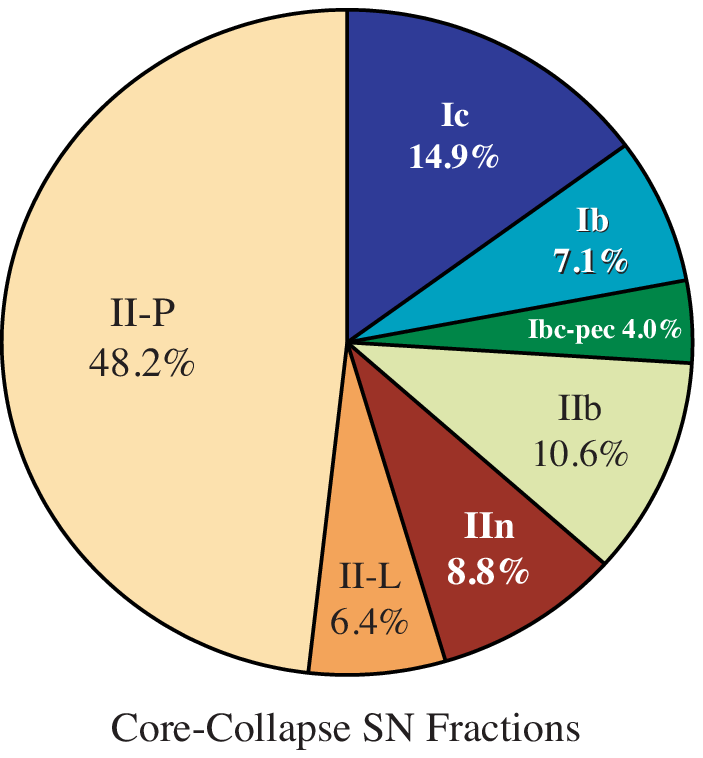}
\end{center}
\caption{Relative fractions of CCSN types in a volume-limited sample
  from LOSS.  This is slightly different from the fractions quoted in
  Paper II, in order to better suit the aim of this paper as explained
  in the text.  The main difference is that we exclude SNe in highly
  inclined galaxies because of extinction effects, and we reorganise
  the class of SNe~Ibc-pec (namely, we moved broad-lined SNe~Ic from
  the ``Ibc-pec'' category to the ``Ic'' group).  }
\label{fig:pie}
\end{figure}

\section{OBSERVED CCSN FRACTIONS}

Figure~\ref{fig:pie} shows a pie chart illustrating the relative
fractions of different types of CCSNe derived from LOSS.  These values
are taken from the volume-limited fractions of all SN types derived in
Paper~II, with the thermonuclear (Type~Ia) explosions subtracted from
the sample.  The relative fractions of the total for CCSNe are listed
in Table 1, and these values are adopted throughout this work.  See
Paper~II for further details on how these numbers are derived from our
survey.  Errors in Table~1 were estimated using a random Poisson
number generator to sample from a list of fake SNe with fractions
corrected for various observing biases, with 10$^6$ realizations.
Paper~II discusses this in more detail.

There are several important points to note here.  This volume-limited
sample of CCSNe excludes most of the so-called ``SN impostors'' (e.g.,
Van Dyk 2010; Smith et al.\ 2010, in preparation), which appear as
relatively faint SNe~IIn that are often discovered by KAIT.  If we had
included them, the fraction of SNe~IIn would be significantly higher;
note that even without the SN impostors, however, our relative
fraction of SNe~IIn is higher than in previous studies (Cappellaro et 
al.\ 1999; Smartt 2009).  The criteria for excluding an individual SN
impostor are admittedly somewhat subjective, but this is a necessary
step since the diversity and potential overlap of SNe~IIn and massive
star eruptions are not fully understood yet.  Generally, if an object
has a peak absolute $R$ or unfiltered magnitude brighter than $-15$ and
has line widths indicating expansion speeds faster than about 1000 km
s$^{-1}$, we include it as a real SN~IIn.  Less luminous and slower
objects are considered impostors and are excluded.

Unlike previous studies, we include a category called ``SNe~Ibc-pec''
(peculiar; see Paper II).  This category was necessary to introduce in
Paper~II because some SN~Ibc vary significantly from the template
light curves used to derive the control times for SNe~Ib and Ic.  As
such, the ``Ibc-pec'' category in Paper II includes some broad-lined
SNe~Ic such as SN~2002ap that are clearly SNe~Ic.  We have moved these
to the SN~Ic category for the purpose of this paper, since they
clearly correspond to massive stars that have fully shed their H and
He envelopes.  This has a small effect on the overall statistics,
because broad-lined SNe~Ic are very rare in our sample, contributing
only 1--2\% of all CCSNe.  This is in agreement with the recent study
of Arcavi et al.\ (2010), who find that broad-lined SNe~Ic contribute
only 1.8\% of CCSNe in large galaxies.  It is noteworthy, however,
that Arcavi et al.\ (2010) find broad-lined SNe~Ic to be much more
common ($\sim$13\% of CCSNe) in low-metallicity dwarf host galaxies.
We also exclude SNe occurring in highly inclined galaxies, where dust
obscuration may introduce statistical problems that are difficult to
correct.  As a result of these minor adjustments, made because our
goal of investigating implications for massive-star evolution is
different from the goal of deriving relative rates and correcting for
observational biases, the relative fractions of various SN types in
Table~1 and Figure~\ref{fig:pie} differ slightly from the results in
Paper~II.


In quoting fractions of various SN types, we ignore metallicity, 
galaxy class, and other properties, although we are cognizant of the
importance of these properties and consider them in our discussion
below.  The galaxies included in the LOSS survey span a range of
luminosity, with most of the CCSN hosts corresponding roughly to
metallicities of 0.5--2 $Z_{\odot}$ (Garnett 2002; the LOSS galaxy
sample spans a range of $M_K$ from about $-$20 to $-$26 mag, but most
of the CCSN hosts are in the range $-$22 to $-$25 mag; see Paper~II).
We note some trends in Paper~II, such as the fact that SNe~IIn appear
to prefer lower luminosity spirals, whereas SNe~Ibc seem to prefer
large galaxies and therefore higher metallicity, consistent with previous
studies (Prantzos \& Boissier 2003; Prieto et al.\ 2008; Boissier \& 
Prantzos 2009).  LOSS is
biased against very faint dwarf galaxies, since larger galaxies with
potentially more SNe were targeted to yield a richer harvest of SNe. However,
low-luminosity galaxies seem to have more than their expected share of 
star formation per unit mass, and probably contribute 5--20\% of the local
star formation (Young et al.\ 2008).  If unusually luminous SNe~IIn
and II-L favour such low-luminosity galaxies, as some recent studies
may imply (Smith et al.\ 2008; Miller et al.\ 2009; Quimby et al.\ 
2009), then this may slightly raise the relative fractions of SNe~IIn
and II-L compared to our study.  Recently commissioned untargeted
surveys can help constrain this contribution (see Arcavi et al.\ 2010,
as noted above regarding broad-lined SNe~Ic in dwarf hosts).


Our volume-limited survey within 60 Mpc includes 80 CCSNe, compared to
the heterogeneous volume-limited study of 92 CCSNe within 28 Mpc
summarised by Smartt (2009).  However, because the LOSS survey was
conducted with the same telescope in a systematic way, we are able to
make proper corrections for the observing biases, as Paper II
describes in detail.  We also have much more complete spectroscopic
follow-up observations and we monitor the photometric evolution of the
SNe we discovered, which particularly affects the relative fractions
of SN~II-P vs.\ II-L, IIn, and IIb, all of which are sometimes called
simply ``Type II'' in initial reports.  Thus, samples of SNe using
identifications from initial reports are often unreliable or
unspecific, but our study resolves this issue because our more
extensive photometric and spectroscopic follow-up observations allow
us to more reliably place the SNe in subclasses.  Consequently, our
observed fractions of CCSN types differ from those of previous studies
in a few key respects.  The main differences compared to SN fractions
listed in various studies reviewed by Smartt (2009) are as follows.

(1) We find a lower SN~II-P fraction of only $\sim$48 \%, in contrast
to larger values of 59\% in previous studies, although some of these 
did not differentiate among SN~II subtypes.  This impacts the ``RSG
problem'' as discussed below.

(2) We find correspondingly larger fractions of SNe~II-L, IIn, and IIb
compared with Smartt et al.\ (2009).  This mostly reflects our
spectroscopic and photometric follow-up observations mentioned above.

(3) We find a larger fraction of SNe~Ibc than Cappellaro et al.
(1997), although similar to other estimates (van den Bergh et al.\ 1987;
Prantzos \& Boissier 2003; Prieto et al. 2008; Boissier \& Prantzos
2009; Smartt et al.\ 2009).
The number ratio of SNe~Ibc to all SNe~II that we measure is
$N_{\rm Ibc}/N_{\rm II}$ = 0.35, whereas Cappellaro et al.\ found a
value for $N_{\rm Ibc}/N_{\rm II}$ of only 0.29.  Prieto et al.\
(2008)\footnote{Note, however, that Prieto et al.\ used the Sternberg
  Astronomical Institute (SAI) SN catalog, which is a heterogeneous
  sample with unknown systematic biases.}  noted that $N_{\rm Ibc}/N_{\rm
  II}$ = 0.27 in the full sample they considered, but they also found a
metallicity dependence, with higher values comparable to ours at
around solar metallicity.  The high ratio we find is the crux of the
``WR problem'' that we discuss herein.

\section{THE IMF AND PROGENITOR MASSES}

The IMF describes the relative number of stars as a function of
initial mass, $N(m)$, and within a given mass range this dictates the
distribution of initial masses for progenitors of SNe.  We adopt a
simple approximation of the IMF as a single power law and exponent
$\gamma$ given by

\begin{equation}
N(m) = C m^{\gamma} ,
\end{equation}

\noindent where $C$ is a constant.  To understand the implications of
SN rates for massive stars, we investigate the IMF within a mass range
bounded by the lowest initial mass that results in a CCSN,
$M_{\rm SN}$, and extending up to the upper mass limit for the initial
masses of stars.  One expects $M_{\rm SN}$ to be around 8 M$_{\odot}$,
but there are uncertainties involved, as discussed further below.  We
take the upper limit to initial masses to be 150 M$_{\odot}$ (Figer
2005), although this choice has little effect on our analysis because
the most massive stars are so rare in the local universe (all the
stars from 100 to 150 M$_{\odot}$ make up less than 2\% of the
population, comparable to our uncertainties).  A handy quantity is
$F_m$, which we define as the fraction of all CCSNe contributed by
stars with initial mass $m$ or higher, up to 150~M$_{\odot}$.  For an
unbroken power-law IMF, this is given by

\begin{eqnarray}
F_m & = & \int_m^{150} N(m') dm' \\
    & = & \frac{[1-(m/150)^{\gamma+1}]}{[1-(M_{\rm SN}/150)^{\gamma+1}]} ,
\end{eqnarray}

\noindent where $\gamma$ = $-$2.35 for a standard Salpeter (1955) mass
function (note that this differs from the logarithmic form that is
sometimes used, where Salpeter corresponds to $\Gamma = -$1.35).
Bastian et al.\ (2010) have provided a recent review of the literature
on possible variations in the IMF, and conclude that there is no clear
evidence that the IMF varies strongly in the modern universe.
Clearly, $1-F_m$ is the cumulative fraction contributed by stars
between the lower bound ($M_{\rm SN}$) and $m$.  This assumes that SN
progenitors occupy the full mass range from $M_{\rm SN}$ to 150
M$_{\odot}$, with no large mass interval where stars consistently
collapse directly to a black hole without any visual display (Fryer
1999); the latter remains a possibility, and implications are
discussed later.

\begin{figure}
\begin{center}
\includegraphics[width=3.0in]{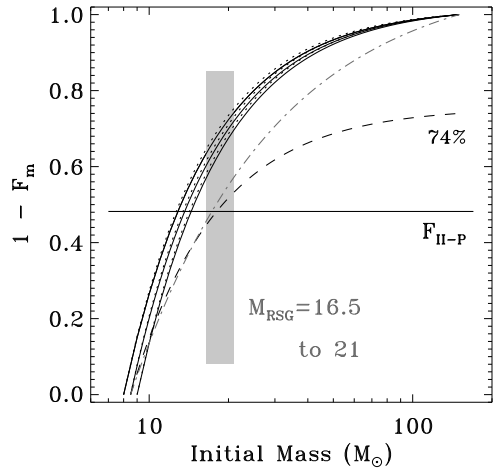}
\end{center}
\caption{The quantity $1 - F_m$.  This is a cumulative distribution
  function beginning at the bottom of the mass range for CCSNe
  ($M_{\rm SN}$), showing the fraction of CCSNe contributed by stars
  in the mass range from $M_{\rm SN}$ up to $m$. The three solid black
  curves are for three example values of $M_{\rm SN}$ = 8.0, 8.5, and
  9.0 M$_{\odot}$ using a Salpeter slope of $\gamma$ = $-$2.35.  The
  dotted curves are for $\gamma$ = $-$2.4, whereas the grey dot-dashed
  curve illustrates the hypothetical top-heavy case of $\gamma$ =
  --1.8 (see text).  The long-dashed curve labeled ``74\%'' shows the
  same function for $M_{\rm SN}$ = 8.5 M$_{\odot}$ and $\gamma$ =
  --2.35, but it excludes 26\% of the total number (26\% is the sum of
  the fractions of all SNe~Ibc), assuming that they follow a different
  evolutionary path in close binaries over the full mass range
  considered; this possibility is discussed later in \S 4.2 and 4.3.
  The gray box denotes the range of uncertainty in the upper bound to
  RSG progenitors of SNe~II-P, based on the properties of progenitors
  detected so far (Smartt et al.\ 2008).  The horizontal line is the
  observed fraction of SNe~II-P.}
\label{fig:fm2}
\end{figure}

\begin{figure}
\begin{center}
\includegraphics[width=3.0in]{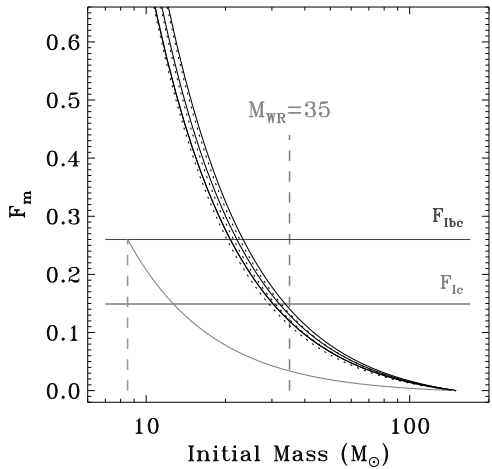}
\end{center}
\caption{The quantity $F_m$ (Eqn.\ 3).  This is similar to
  Figure~\ref{fig:fm2}, but with a cumulative distribution function
  beginning at the {\it upper} mass limit and working down, showing
  the fraction of stars in the mass range between $m$ and the assumed
  upper mass limit at 150 M$_{\odot}$.  The horizontal lines mark the
  observed fractions of SNe~Ic and the sum of SNe Ib + Ic + Ibc-pec.
  The vertical dashed line at $M_{\rm WR}$ = 35 M$_{\odot}$ marks the
  initial mass above which H-free WR stars are thought to originate,
  inferred from observations.  The grey curve is the same as the solid
  black curve corresponding to $M_{\rm SN}$ = 8.5 M$_{\odot}$, but
  multiplied by 0.26 to mimic the distribution of SNe~Ibc if they were
  evenly distributed across the full mass range.}
\label{fig:fm}
\end{figure}

Figures~\ref{fig:fm2} and \ref{fig:fm} show plots of $1-F_m$ and
$F_m$, respectively, for three different representative values of
$M_{\rm SN}$ = 8.0, 8.5, and 9.0 M$_{\odot}$, as well as for two
different values of $\gamma$ = --2.35 (Salpeter 1955) or --2.4 (e.g.,
Humphreys \& McElroy 1984) for comparison.  One can see that small
variations in $\gamma$ have little effect on the results.
Figure~\ref{fig:fm2} also illustrates a hypothetical case of $\gamma$
= $-$1.8, which is large enough to make a substantial difference (this
is the slope that would be needed to reconcile the disagreement
between the observed fraction of SNe~II-P and the observed mass range
for the corresponding progenitors; see below).  This slope, however,
is more top-heavy than allowed by measurements of local stellar
populations outside of the inner parts of the densest star clusters
(see Bastian et al.\ 2010).

Small differences in the adopted value of $M_{\rm SN}$ can have a
substantial effect, however.  This is due to the fact that lower-mass
stars are so much more numerous in a bottom-heavy IMF, and small
changes in $M_{\rm SN}$ therefore have a disproportionate influence on
the distribution of SN types.  This is relevant in regard to the
still-uncertain lower bound to initial masses that experience Fe core
collapse and those that may suffer less energetic explosions via
electron-capture SNe (ECSNe).  According to Smartt (2009), directly
observed RSG progenitors of normal SNe~II-P extend down to around 8
M$_{\odot}$ and their statistical distribution favours $M_{\rm SN} =
8.0 \pm 1.0$ M$_{\odot}$.  On the other hand, theories for ECSNe
predict that these explosions occur somewhere in the range 8--11
M$_{\odot}$ depending on assumptions about metallicity, mass loss, and
other factors (Nomoto 1984; Woosley et al.\ 2002; Kitaura et al.\
2006; Wanajo et al.\ 2009; Pumo et al.\ 2009).  Theory generally
predicts that if ECSNe occur within this range, they would tend to be
less energetic and fainter than a standard Fe CCSN, releasing
$\sim$10$^{50}$ erg of kinetic energy (instead of $\sim$10$^{51}$ erg)
and producing less $^{56}$Ni than standard CCSNe.  It has been
hypothesised that an ECSN may have given birth to the Crab Nebula
(Davidson et al.\ 1982; Nomoto 1984; Kitaura et al.\ 2006; Wanajo et
al.\ 2009).

A note of caution is that if the corresponding observed visual
displays are indeed much fainter than normal CCSNe, then it is
possible that some of these ECSNe may not be included in the LOSS
sample, since we chose to exclude faint Type~IIn events such as SN
impostors, $\eta$~Car analogs, LBVs, or other peculiar faint
transients in the observed fractions of CCSNe (Paper II).  On the
other hand, if ECSNe do not give rise to these SN impostors, but
appear instead as the relatively faint end of the distribution of
SNe~II-P (objects like SN~2005cs; Pastorello et al.\ 2007), then they
will be included in the LOSS rates as SNe~II-P.  The luminosity
functions in Paper~II reveal an enhancement at the very bottom of the
luminosity range of SNe II-P.  This uncertainty is unfortunate, but
the ECSN phenomenon is not understood sufficiently well to confidently
account for it.  For this reason, Figures~\ref{fig:fm2} and
\ref{fig:fm} show values for $M_{\rm SN}$ of 8.0, 8.5, and 9.0
M$_{\odot}$ and the range of uncertainty that this introduces.

Independent of the questions surrounding ECSN theory, however, an
empirical value of $M_{\rm SN}$ = 8.0$\pm$1.0 M$_{\odot}$ is favoured
by Smartt (2009) based on the distribution of masses for directly
observed SN~II-P progenitors (although one must remember that this
value is model dependent as well, and subject to systematic effects;
see Smartt 2009 for details).  We adopt $M_{\rm SN}$ = 8.5 M$_{\odot}$
for most discussion in this work.  If the ECSN phenomenon occurs above
8.5 M$_{\odot}$, we consider it likely that those ECSNe will be
included among the population of faint SNe~II-P anyway, while those
below could be excluded if they masquerade as faint transients or SN
impostors (e.g., Thompson et al.\ 2009).

\subsection{The RSG Problem}

Red supergiants (RSGs) represent the expected endpoint of
post-main-sequence stellar evolution for the majority of single stars
with initial masses above 8 M$_{\odot}$, and it is straightforward to
associate their extended H-rich envelopes with SNe~II-P --- the most
common type of CCSN.  This has long been expected (e.g., Falk \&
Arnett 1977; Litvinova \& Nadyozhin 1983; Doggett \& Branch 1985; 
Wheeler \& Swartz 1993), but the RSG/II-P connection has received firm
footing in the past decade with the identification of RSGs as the
progenitor stars of several SNe~II-P.  This work has been based on
attempting to locate progenitor stars (or upper limits to them) in
pre-explosion archival data at the same position as the SN (Barth
et al.\ 1996; Van Dyk et al.\ 1999, 2003a, 2003b, 2003c; Smartt et 
al.\ 2001, 2002, 2003, 2004; Li et al.\ 2005, 2006, 2007; Maund 
\& Smartt 2005), and in some cases the RSG disappears after the 
SN has faded.

These multiple progenitor studies have reassured us that RSGs are the
progenitors of SNe~II-P, but what range of initial masses do they
imply?  There are many potential systematic errors involved: masses
derived from progenitor luminosities rely upon model-dependent
evolutionary tracks, and circumstellar dust that may have surrounded
the progenitor could have been vaporised by the SN, causing the extinction
derived toward the SN progenitor --- and therefore its luminosity and
mass --- to be underestimated.  Smartt (2009) has reviewed the recent
literature on the identification of SN~II-P progenitors as RSGs in
pre-explosion data and discussed these systematics.  Altogether,
Smartt (2009) argues that the available collection of SN II-P
progenitor detections and upper limits favours 8.5--16.5 M$_{\odot}$
for the range of initial masses, adopting a normal Salpeter IMF, and
Smartt et al.\ (2009) give an upper limit to initial masses of SN II-P
progenitors of 21 M$_{\odot}$ with 95\% confidence.  The upper limit
in the range of 16.5 to 21 M$_{\odot}$ is shown by the gray shaded
area in Figure~\ref{fig:fm2}.

In our volume-limited sample of SNe, we find that SNe~II-P constitute
about 48\% of CCSNe (Figure~\ref{fig:pie}).  This is a lower fraction
than reported in previous studies (Smartt 2009; Smartt et al.\ 2009;
note that several other previous studies did not explicitly separate
SNe~II-P from II-L or other SNe~II).  Figure~\ref{fig:fm2} compares
this LOSS observed fraction of SNe~II-P, $F_{\rm II-P}$, to the
quantity $1-F_m$ (see Eqn.\ 3), which is the fraction of CCSNe one
expects integrating from the bottom of the CCSN range at $M_{\rm SN}$
= 8.0, 8.5, or 9.0 M$_{\odot}$ up to mass $m$.

From Figure~\ref{fig:fm2} we see that the initial mass range of
8.5--16.5 M$_{\odot}$ over which RSG progenitors of SNe~II-P have been
identified would provide more than enough SNe to account for the
observed fraction of SNe II-P, under the assumption that all stars
within this mass range explode as SNe~II-P.  In fact, stars in the
initial mass range 8.5--16.5 M$_{\odot}$ would constitute roughly 62\%
of all the stars above 8.5 M$_{\odot}$ (for $\gamma$ = $-$2.35) that
undergo core collapse, producing {\it too many} SNe~II-P.  The mass
range 8.5--13.7 M$_{\odot}$ would be sufficient to produce the
observed fraction of SN II-P.

Thus, there is apparently no RSG problem from the ``supply-side''
point of view, in the sense that the observed range of masses for
SN~II-P progenitors supplies a {\it large enough} fraction of CCSNe.
Looking more closely, however, there is a ``demand-side'' problem in
the sense that stars in the initial mass range of 8--17 M$_{\odot}$
--- which are in fact observed to explode as SNe~II-P --- produce {\it
  too many} SNe~II-P compared to the observed fraction of this SN
subtype.  Smartt et al.\ (2009) did not emphasise this discrepancy in
their study, presumably because they concluded that SNe~II-P
constitute a larger fraction ($\sim$59\%) of CCSNe, which would be in
reasonable agreement with the observed mass range within their
uncertainties, compared to our value of $\sim$48\%, which is
discrepant.  A suggestive solution is given by the dashed line in
Figure~\ref{fig:fm2}, which brings the observed fraction of SNe~II-P
and the mass range of detected SN~II-P progenitors into agreement.
This curve is the same as the value of $1-F_m$ shown by the black
curves ($\gamma$ = $-$2.35; $M_{\rm SN}$ = 8.3 M$_{\odot}$), but
multiplied by 74\%.  The motivation for this is that it assumes that
the 74\% of CCSNe that are Type~II are distributed evenly across all
initial masses, and that therefore the 26\% of all SNe that are
SNe~Ibc have some different origin which is also distributed across
all initial masses.  This might be the case, for example, if all
SNe~Ibc arise from RLOF in binary systems.  We return to this question
later.

Note that our comment about the lack of any ``supply-side'' RSG
problem is different from the RSG problem pointed out by Smartt et
al.\ (2009), which has to do with the fact that RSG stars in the
initial mass range $17 < m < 25 \ {\rm M}_{\odot}$ are observed to exist,
yet they appear to be missing from the population of nearby SN~II-P
progenitors detected in pre-explosion data.  Smartt et al.\ (2009)
hypothesised that these missing progenitor stars may collapse directly
to black holes without producing successful SNe.  Another possible
solution to this discrepancy, however, is that RSGs in this upper mass
range continue to evolve into other types of progenitor stars before
core collapse, such as yellow supergiants (YSGs), blue supergiants
(BSGs), low-luminosity LBVs, or Wolf-Rayet (WR) stars, producing SNe
of Types II-L, IIb, IIn, Ib, or Ic.  Smartt et al.\ (2009) mentioned
this hypothesis but disfavoured it, in part because the number of
SNe~IIb + II-L + IIn was not enough to make up for the missing
population of RSGs, plus other reasons concerning the inferred masses
for LBVs and progenitors of events like SN~1993J and SN~1980K.
However, we find that these arguments rely on unreliable assumptions
and that they provide no compelling argument against the idea that
RSGs in the initial mass range 17--25 M$_{\odot}$ may continue to
evolve before exploding.

Furthermore, in the volume-limited sample from LOSS, we find that SNe
II-L, IIb, and IIn make up a larger fraction of the total SN~II group,
and SNe II-P have a lower fraction, compared to the study of Smartt et
al.\ (2009).  With this LOSS sample, we find that there are plenty of
remaining SNe~II besides SNe~II-P to account for SNe resulting
(eventually) from RSGs known to occupy the higher mass ranges above 17
M$_{\odot}$.  Another objection stems from the assumption by Smartt et
al.\ (2009) that LBVs (the likely progenitors of SNe~IIn) arise
exclusively from stars with initial masses above 40 M$_{\odot}$, but
there is also a well-known population of lower-luminosity LBVs that
are thought to be stars in a post-RSG phase with initial masses of
20--40 M$_{\odot}$ (see Smith et al.\ 2004).  In addition to LBVs,
Smith et al.\ (2009) noted that the most extreme class of RSGs with
high mass-loss rates and initial masses of 25--35 M$_{\odot}$ could
give rise to the lower-luminosity SNe~IIn.  Thus, these considerations
alleviate two key objections to the idea that 17--25 M$_{\odot}$ stars
produce other types of SNe that are not Type II-P.

In fact, there is growing empirical evidence that this is indeed the
case, supported by direct detections of progenitor stars of SNe II-L,
IIb, and IIn (and II-pec).  SN~2009kr is the first luminous SN~II-L to
have a progenitor star identified in pre-explosion images (Elias-Rosa
et al.\ 2010b; Fraser et al.\ 2010), and it appears to be a YSG.
Elias-Rosa et al.\ (2010b) estimate a likely initial mass for the YSG
progenitor star of 18--24 M$_{\odot}$, and infer that it may bridge a
gap in progenitor mass between SNe~II-P and the more massive LBV
progenitors of SNe IIn (see below).  SN~2009hd in M66 also had a
Type~II-L spectrum, for which Elias-Rosa et al.\ (2010c) have
identified another likely YSG progenitor, suggesting an initial mass
in the range 20--25 M$_{\odot}$.  SN~2008cn is yet another possible
YSG progenitor of a luminous SN~II-P (Elias-Rosa
et al.\ 2010a), although the large distance to this SN makes the
progenitor identification less secure.

Of course, the first SN to have a progenitor identified in
pre-explosion data was SN~1987A, whose classification was Type II-pec,
and which was inferred to have an $\sim$18 M$_{\odot}$ BSG progenitor
that was in a post-RSG phase (see Arnett 1987; Arnett et al.\
1989). The progenitor of the SN~IIb 1993J was inferred to be a $M_0 \
\approx \ 15 \ {\rm M}_{\odot}$ K-type RSG with a large radius but
small H envelope mass (Aldering et al.\ 1994; Filippenko et al.\ 1994;
Van Dyk et al.\ 2002; Maund et al.\ 2004).  In both cases, binary
evolution was invoked to explain the status of the progenitors at the
time of core collapse (Nomoto et al.\ 1993; Podsiadlowski et al.\ 
1993; Aldering et al.\ 1994; Woosley et al.\ 1994).  Lastly, so far
only one SN~IIn (SN~2005gl) has a progenitor star identified in
pre-explosion data, and it was a massive LBV (Gal-Yam \& Leonard
2009).

Collectively, these results argue that RSGs in the range of masses
above the observed range for SN~II-P progenitors may indeed continue
to evolve after the RSG phase due to further mass loss (in either
single- or binary-star evolution), to produce other types of SNe.
This relieves the RSG problem proposed by Smartt et al.\ (2009), and
removes the empirical motivation for inferring that massive stars in
some mass range collapse directly to a black hole (BH) without a
visible SN display.  In fact, we find that the latter inference would
introduce other problems that are at odds with the observed fractions
of CCSNe, as discussed further below.

\subsection{The WR Problem}

Unlike the detected progenitors of SNe~II-P, II-L, IIb, and IIn, the
progenitor stars that have shed their H envelopes to make SNe~Ibc have
never yet been identified directly in pre-explosion data.\footnote{One
  putative exception is SN~2010O, for which Nelemans et al.\ (2010)
  detect a variable X-ray source at the SN position in pre-explosion
  data.  Nelemans et al.\ (2010) claim that this may have been a
  WR/black hole binary system, where SN~2010O was the second SN in the
  system when a WN star produced the observed SN~Ib.  The WN star
  itself was not detected, however.}  The known stars that most
naturally fit the bill for progenitors of SNe~Ib and Ic are the
Wolf-Rayet (WR) stars of the WN and WC subclasses, respectively,
because of their relatively H-free surface composition and their small
stellar radii (e.g., Woosley \& Bloom 2006).  The distribution of WR
stars in galaxies appears marginally consistent with that of SNe~Ibc,
although this depends on metallicity (Leloudas et al.\
2010).\footnote{We do not include the group of luminous H-rich
  late-type WN stars, or WNH stars (Smith \& Conti 2008), which are
  probably still in core-H burning and are more like O-type stars with
  enhanced winds.}

It is straightforward to expect that the WC subclass would explode to
produce SNe~Ic, but it is not so clear if the WN subclass explodes as
SNe~Ib, or if instead the WN stars should continue to evolve by virtue
of their own mass loss to become WC stars before exploding as SNe Ic.
Evidence that some WN evolve to WC is that the WN/WC ratio is $\sim$1
in the Milky Way and higher at lower metallicity, whereas from LOSS we
find that N$_{\rm Ib}$/N$_{\rm Ic}$ is only $\la$0.5.  This interplay
may be luminosity and metallicity dependent (as discussed further
below), and comparisons of WR and SN~Ibc positions in galaxies give
mixed results (Leloudas et al.\ 2010).  The fact that no normal SN~Ib
or Ic has an identified progenitor star\footnote{A pre-explosion
  source was identified at the position of the famous object
  SN~2006jc, but this is a highly unusual case.  The pre-explosion
  object was seen only as a transient source in a brief eruptive phase
  2 yr before the SN, and the subsequent SN was a very unusual event,
  probably an underlying SN~Ic whose shock overtook a dense He-rich
  circumstellar shell to produce a SN~Ibn with bright, relatively
  narrow He~{\sc i} lines in the spectrum (see Pastorello et al.\
  2007; Foley et al.\ 2007; Smith et al.\ 2008a).  Nevertheless, the
  $\sim$10$^3$ km s$^{-1}$ speed of the pre-shock CSM seems consistent
  with a compact WR-like progenitor.}  makes the identification of
luminous WR stars as the only progenitors of both SNe~Ib and Ic
uncertain.


Standard single-star evolution models (e.g., Meynet et al.\ 1994;
Heger et al.\ 2003) predict that strong line-driven stellar winds at
high luminosity will cause stars more massive than some threshold
mass to completely shed their H envelopes.  This leaves He cores that
are observed as WR stars (e.g., Conti 1976), and which should explode
to make SNe~Ibc.  For convenience, we define $M_{\rm Ibc}$ as the initial
mass dictated by the observed fraction of SNe~Ibc, above which all
progenitors have fully shed their H envelopes before core collapse.
From Figure~\ref{fig:fm}, we find that SN statistics from LOSS
show that $M_{\rm Ibc} \approx 22 {\rm M}_{\odot}$.

Similarly, we define $M_{\rm WR}$ as the initial stellar mass above
which a massive star is expected to shed its H envelope.  If standard
single-star evolution applies, then we should find $M_{\rm Ibc} =
M_{\rm WR}$.  However, standard single-star evolutionary models such
as those by Heger et al.\ (2003) predict a much higher value of
$M_{\rm WR}$ = 34 M$_{\odot}$ at solar metallicity, and they suggest
that $M_{\rm WR}$ rises to even higher initial masses at lower
metallicity due to the strong metallicity dependence of line-driven
winds that are assumed to dominate.  A problem recognised in recent
years is that these single-star evolutionary models have used
empirical prescriptions for mass-loss rates that are now known to be
{\it far too high} by factors of 3--10 compared to observed mass-loss
rates, as noted in \S 1.  Using more realistic wind mass-loss rates
would change the predictions significantly, such that single-star
evolution would not be able to account for the population of WR stars
or SNe~Ibc, even at solar metallicity.\footnote{Models with rotation
  (e.g., Meynet et al.\ 2008) have been proposed to yield lower values
  of $M_{\rm WR}$ as low as $\sim$25 M$_{\odot}$ that are in better
  agreement with $M_{\rm Ibc}$, but this is because they have even
  higher mass-loss rates, violating observational constraints even
  more severely; this makes rotation an unlikely solution.}  Smith \&
Owocki (2006) have discussed this, pointing out that giant LBV-like
eruptions may provide a way to make up the deficit, but the mass range
over which this applies is uncertain; eruptive mass loss is probably
dominant in only the most massive stars.  Observations of WR stars
associated with star clusters suggest a value of $M_{\rm WR}$ around
roughly 35--40 M$_{\odot}$ for most WR stars in the Milky Way (Schilde
\& Meader 1984; Massey et al.\ 1995, 2001; Crowther et al.\ 2006;
Crowther 2007; Massey 2003; Humphreys et al.\ 1985).  We therefore
adopt $M_{\rm WR}$ = 35 M$_{\odot}$ for the majority of WR stars at
roughly solar metallicity.\footnote{There is also evidence that some
  lower-luminosity early-type WN stars may originate from lower
  initial masses down to $\sim$25 M$_{\odot}$ (Crowther 2007), but
  stellar winds at solar metallicity (and probably even LBV eruptions
  as well) are insufficient to strip their H envelopes.  In close
  binary evolution, however, complete removal of the H envelope can
  occur at much lower initial masses down to $\sim$15 M$_{\odot}$ or
  less (e.g., Eldridge et al.\ 2008; Podsiadlowski et al.\ 1992).  We
  will return to this later.}

Figure~\ref{fig:fm} highlights a serious problem with assigning
classical WR stars as the exclusive progenitors of SNe~Ibc.  Namely,
the fraction of all stars experiencing CCSN above $M_{\rm WR}$ = 35
M$_{\odot}$ only accounts for about half the number needed for the
observed fraction of SNe~Ibc.  To account for all SNe~Ibc in this
simple prescription --- where more luminous stars have higher
mass-loss rates and therefore become WR stars and SNe~Ibc by virtue of
their own mass loss --- would require that SN~Ibc progenitor stars
extend from the upper mass limit down to around 22 M$_{\odot}$.  In
other words, the WR problem can be stated simply as

\begin{equation}
M_{\rm WR} >> M_{\rm Ibc}.
\end{equation}

According to Figure~\ref{fig:fm}, roughly half the SN~Ibc population
must originate from stars that are less massive than initial masses
corresponding to the observed or theoretically expected population of
WR stars. There are several possible solutions to this problem. (1)
The WR phase for many lower-mass stars is not observed because it is
extremely short-lived, perhaps because eruptions in late evolutionary
phases remove the remaining H envelope even down to lower masses than
we normally associate with LBVs.  In this case, however, some SNe~Ibc
should show signs of interaction with H-rich CSM at late times,
because that H must have been shed recently, while only a few do.
SN~2001em is one example (Chugai \& Chevalier 2006; Schinzel et al.\
2009; Van Dyk et al.\ 2009), but perhaps there are more where the CSM
interaction is missed at very late times.  (2) Alternatively, the
population of H-free stars that correspond to the progenitors of
almost half of SNe~Ibc may be underluminous because of significant
mass loss in binary RLOF.  If underluminous, their radiation-driven
winds --- and hence, their emission-line spectra --- may be weak and
so they are not discovered or identified observationally as classical
WR stars. These may be hidden by brighter companion stars in binary
systems (i.e. the overluminous mass gainers), making them more
difficult to observe.  Smartt (2009) mentioned this as a potential
explanation for the lack of any detection of SN~Ibc progenitors so
far.  The idea that RLOF dominates the population of SN~Ibc
progenitors was suggested long ago (e.g., Filippenko 1991; Branch,
Nomoto, \& Filippenko 1991; Podsiadlowski et al.\ 1992) but has been
hard to confirm.  We find, as discussed further below, that {\it this
  is the likely origin of at least half and possibly most SNe~Ibc}.


This is not to say that the expectations of single-star evolution are
completely irrelevant.  While binary RLOF may be largely independent
of metallicity and initial mass, stellar winds may still play an
important role.  RLOF in binaries provides a likely way to strip the H
envelopes at any metallicity, but it is less likely to strip the He
envelope except for the shortest-period systems.  The same goes for
shedding the H envelope via giant LBV-like eruptions, which may also
be insensitive to metallicity (Smith \& Owocki 2006).  However, the
{\it subsequent} evolution of the stripped He core --- from one with a
small residual H mass to a H-free and He-rich surface, and eventually
toward removal of the He layer as well --- can be accomplished by the
line-driven wind of the WR star itself, {\it which does depend
  strongly on both metallicity and luminosity (and therefore initial
  mass)}.  This is supported by the observation that the ratio of WN
stars to WC stars varies from $\sim$1 in the Milky Way to 5 and 10 in
the LMC and SMC, respectively (e.g., Crowther 2009). Even as binaries,
some of the SNe~Ibc -- in particular the SNe~Ic -- may therefore
appear to obey expected trends of single-star evolution, where the
most luminous and higher metallicity stars are more able to shed their
He envelopes via radiation-driven winds or eruptions, leading to WC
stars and SNe~Ic.  This may explain why studies of the positions of WR
stars and SNe in their host galaxies find that WC stars and SNe~Ic
seem to imply higher initial mass and higher metallicity environments
(Kelly et al.\ 2008; Anderson \& James 2009; Leloudas et al.\ 2010; 
see also Papers I, II, and III), even if binary evolution or LBV
eruptions dominate the removal of the H envelope.  We emphasise that
it will be quite important in future studies to distinguish between
SNe~Ib and Ic while studying SN statistics as functions of metallicity
and redshift.  By the same token, it may be important to distinguish
among subtypes of SNe~II, as discussed next.

\begin{figure*}
\begin{center}
\includegraphics[width=4.6in]{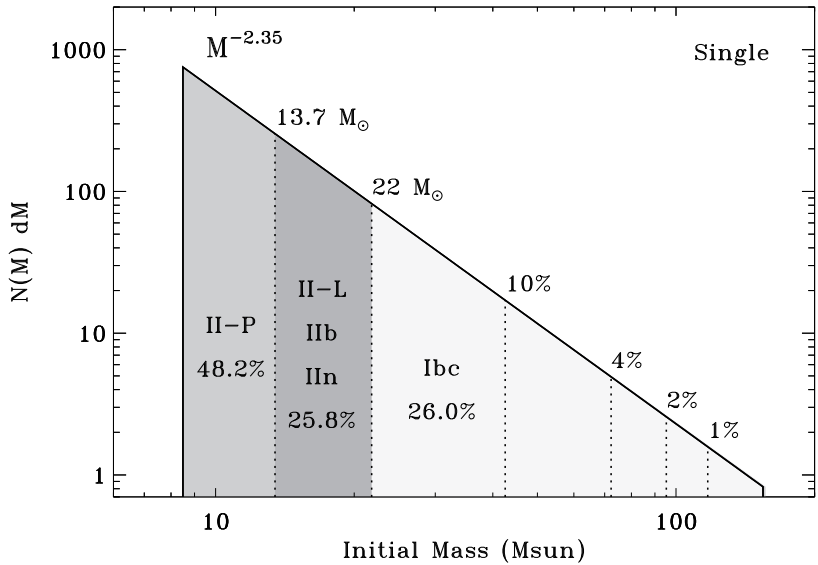}
\end{center}
\caption{Mass ranges implied by the observed fractions of SN types
  in a standard single-star evolutionary framework, where higher
  initial masses lead to higher mass-loss rates, and consequently
  greater stripping of the H and He envelopes.  $M_{\rm Ibc} \approx 22\,
  {\rm M}_{\odot}$ is the dividing point above which stars must fully shed
  their H envelopes in this scenario.}
\label{fig:imfS}
\end{figure*}

\subsection{The LBV Problem}

At odds with the standard scenario for the formation of WR stars as
the descendants of the most massive stars with $M_0 \ga 35 \
{\rm M}_{\odot}$ is the uncertain fate of LBVs and their connections to SNe.
In this standard scenario (e.g., Conti 1976), winds of O-type stars on
the main sequence shed much of the H envelope, leaving a very brief
($\sim$10$^4$ yr) transitional LBV stage at the end of core-H burning
that finishes the job of forming H-free WR stars.  Recent work (Bouret
et al.\ 2005; Fullerton et al.\ 2006; Puls et al.\ 2006) has
demonstrated that O-star winds are clumped and that their mass-loss
rates are too weak, so it appears likely that LBV giant eruptions must
dominate this mass loss if WR stars are to form via single-star
evolution (Smith \& Owocki 2006).  If these eruptions are not strong
enough, the star will fail to shed much of its H envelope before core
collapse, producing a SN~IIn (Smith \& Owocki 2006).

In fact, recent studies have provided mounting evidence that some LBVs
explode as SNe before the stars are able to fully shed their H
envelopes.  Luminous SNe~IIn, which are thought to be powered by shock
interaction with dense CSM, require large masses of material ejected
in sudden eruptions that occur within decades before core collapse, in
some cases as high as 10--20 M$_{\odot}$ (Chugai et al.\ 2004; 
Smith et al.\ 2007, 2008b, 2010; Woosley et al.\ 2007; Smith 2008).
The large CSM masses for luminous events like SN~2006tf and SN~2006gy
require very massive progenitor stars to account for the mass budget,
since the large ejecta mass corresponds only to the H-rich envelope
ejected just before core collapse (i.e., the true initial mass of the
star also includes the He core and any mass shed during the star's
lifetime).  One of the hypotheses for the pre-SN mass ejections of
SNe~IIn is that they suffered pulsational pair instability ejections
before core collapse, in which case {\it very} massive progenitor
stars with $M_0 \ge 95\, {\rm M}_{\odot}$ are needed (Heger et al.\ 2003).

There are other, anecdotal signs of a link between LBVs and SNe IIn as
well, having to do with their wind speeds, absorption profiles, and
circumstellar nebulae (Kotak \& Vink 2006; Smith 2007; Trundle et al.\ 
2008).  Much more directly, Gal-Yam \& Leonard (2009) showed that the
LBV-like progenitor of the SN~IIn 2005gl subsequently disappeared,
providing a strong case that LBVs do in some cases explode as SNe~IIn,
despite that fact that no contemporary stellar evolution models
predict this.  Gal-Yam \& Leonard inferred a high initial mass of
$\ga$50 M$_{\odot}$ for the progenitor of SN~2005gl.  The ``LBV
problem,'' then, is the fact that LBVs or some other very massive,
unstable H-rich stars explode as SNe~IIn, even though current models
expect very massive stars to shed their H envelopes.

If SNe~IIn truly arise from massive LBV progenitors, exactly what
ranges of initial mass are required?  How can we divide the IMF such
that very massive progenitor stars can yield {\it both} SNe~IIn and
the SNe~Ic that are supposed to come from the WC descendants of very
massive stars?  What scenarios are consistent with the observed
fractions of various types of CCSNe?  We investigate this problem
next.


\section{HYPOTHETICAL SCENARIOS FOR DIVIDING THE IMF}

Given the problems and complications between progenitor scenarios
expressed in the previous section, we now address the problem from a
simpler empirical point of view.  Here we ask how one can subdivide
the IMF of massive stars in a way that is consistent with the
fractions of various CCSN types observed in LOSS, while also meeting
requirements imposed by our knowledge of the likely progenitor stars.
For simplicity, in all cases we adopt a Salpeter IMF within the mass
range bounded by the lowest initial mass for which SNe occur, assumed
to be $M_{\rm SN}$ = 8.5 M$_{\odot}$, up to the proposed upper mass
limit for initial masses at 150 M$_{\odot}$ (Figer 2005).

This is meant to be exploratory and demonstrative, rather than
definitive.  We consider extreme hypotheses such as one where all
massive stars obey expectations of single-star evolution (e.g., Heger
et al.\ 2003), and alternatively, where all stripped-envelope SNe
arise from binary RLOF (e.g., Filippenko 1991; Podsiadlowski et al.\
1992), and we evaluate merits and drawbacks of each.  We also mention
a compromise ``hybrid'' scenario.  Our analysis is intended to guide
intuition in future studies, and to provide tests for
single/binary-star population synthesis models.

\subsection{Dominated by Standard Single-Star Evolution}

We first consider the familiar hypothesis that at a given metallicity,
increasingly more massive and more luminous stars have monotonically
increasing mass-loss rates, such that higher initial masses invariably
lead to greater stripping of the H and He envelopes.  It is
essentially a hypothesis that single-star mass loss dominates over
close binary interactions in stripping a massive star's envelope,
thereby determining the distribution of SN types.  This is widely
considered to be the ``standard'' view of mass loss connecting stellar
initial masses to their ultimate fates as a function of metallicity (e.g.,
Heger et al.\ 2003).  In this picture, the most massive stars fully
shed their H envelopes by virtue of their own strong winds or LBV-like
eruptions to produce SNe~Ibc.  At intermediate masses, stars do not
fully shed their H envelopes, instead producing SNe~IIn, IIb, and
II-L, depending on how much H mass was lost, and how recently this
occurred (i.e., the density of the immediate CSM).  The lowest mass
range corresponds to RSGs that do not shed their H envelopes and
produce SNe~II-P.

Figure~\ref{fig:imfS} shows how the IMF would need to be subdivided in
this hypothetical single-star framework, dictated by the observed
fractions of various SNe types determined by LOSS
(Figure~\ref{fig:pie}).  Figure~\ref{fig:imfS} is largely a more
succinct restatement, in graphical form, of the discussion above
concerning inconsistencies with RSGs, WR stars, and LBV progenitors.


While in principle this scenario is consistent with the qualitative
expectation that more massive stars have stronger pre-SN mass loss, it
also comes with many inconsistencies, and conflicts with several
observational constraints on the likely progenitors of various SN
types.  Some obvious problems evident from Figure~\ref{fig:imfS} are
the following:

(1)  The mass range occupied by the observed fraction of SNe~II-P
(8.5--13.7 M$_{\odot}$) is too small compared to the directly observed
mass range of SNe~II-P progenitors, 8.5--17 M$_{\odot}$ or more
(Smartt 2009).  In other words, if all stars in the range 8.5--17
M$_{\odot}$ produced SNe~II-P, then the fraction of CCSNe that are
II-P would be much higher than observed.

(2)  This scenario contradicts the observational indication that some
SNe~IIn have very massive LBV-like progenitors, as discussed above.
A few SNe~IIn appear to have progenitors with initial masses of at
least 50--80 M$_{\odot}$, whereas Figure~\ref{fig:imfS} requires that
no stars above 22 M$_{\odot}$ retain any H envelopes at core collapse.

(3) Similarly, recent identifications of yellow supergiants as
SNe~II-L progenitors place them at the upper extreme of the range
allowed for SNe~II, or even above 22 M$_{\odot}$.  Masses inferred for
the SNe~II-L 2009kr and 2009hd are 18--24 M$_{\odot}$ and 20--25
M$_{\odot}$, respectively (Elias-Rosa et al.\ 2010b; 2010c).

(4) Most importantly, there are far too many observed SNe~Ibc,
requiring that all stars above 22 M$_{\odot}$ completely shed their H
envelopes, whereas the expected value $M_{\rm WR}$ is roughly 35
M$_{\odot}$.  Even at solar metallicity, stars below $M_{\rm WR}$ do
not have sufficiently high mass-loss rates to shed their H envelopes
--- certainly not through metallicity line-driven winds or RSG winds,
and probably not through LBV eruptions either.  The known initial mass
range for most nearby WR stars accounts for only half the SN~Ibc
population.


We regard this disagreement as strong evidence that standard
single-star evolution with mass loss simply cannot account for the
observed distribution of SNe types, and that binary RLOF is therefore
needed to account for at least half of the SN~Ib/Ic population,
possibly most of it (this is the next case discussed below).

Introducing the hypothesis that some stars collapse directly to BHs
without making a visible SN does not help.  If we assume that the most
massive stars collapse to BHs (with $>$40 M$_{\odot}$, for example;
Fryer 1999), then it pushes the dividing mass between SNe~II and
SNe~Ibc to even lower values, making the problem worse.  It would also
push the upper mass for SN~II-P progenitors even lower, causing an
even worse discrepancy with direct progenitor mass estimates.  A
partial solution relying on BHs would require a finely tuned or
carefully chosen set of intermediate mass ranges for BHs, but it is
still unsatisfying (i.e., assuming that stars of, say, 20-30
M$_{\odot}$ initial mass collapse to BHs could bring the mass range of
SNe~Ibc into better agreement with $M_{\rm WR}$, but it would worsen
the problem in points 1-3 above).  Direct SN-less collapse to a BH may
nevertheless be a possibility.  Better constraints on the
disappearance of stars without SNe are needed (e.g., Kochanek et al.\
2008).

\begin{figure*}
\begin{center}
\includegraphics[width=4.6in]{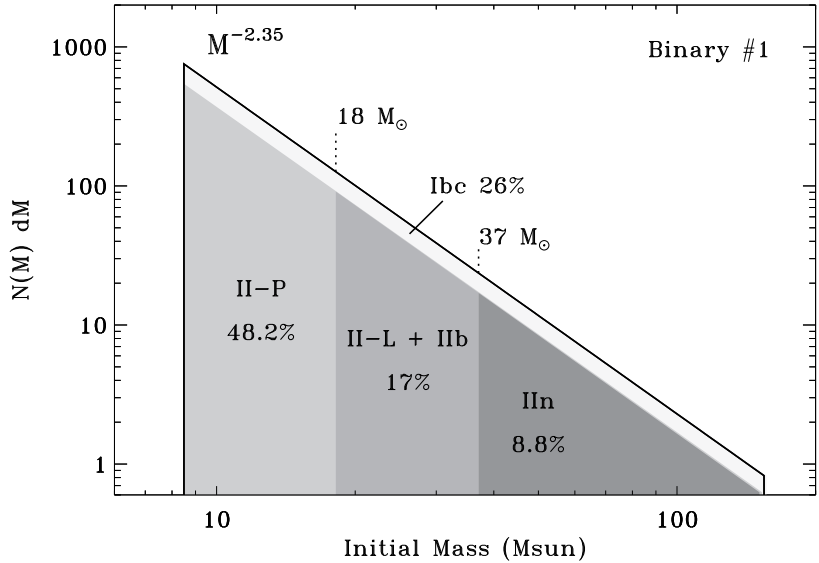}
\end{center}
\caption{Same as Figure~\ref{fig:imfS}, but now assuming that close
  binary evolution and RLOF is a necessary ingredient to explain the
  loss of the H envelope for all SNe~Ibc.  In this ``Binary \#1''
  scenario, the fraction of all massive stars that lose their H
  envelopes in this way is dictated to be 26\%, which is the same as
  the observed fraction of SNe~Ibc.  For simplicity, these binaries
  are divided equally among all initial masses; consequently, the
  remaining stars that fail to shed their H envelopes (all SNe~II) are
  redistributed across the full range of initial masses as well,
  following expectations that more massive stars have higher mass-loss
  rates.}
\label{fig:imfB1}
\end{figure*}

\subsection{Dominated by Close Binaries \#1}

An alternative to single-star mass loss is that mass ejection or mass
transfer via RLOF in interacting binaries plays a dominant role in
stripping away the H envelope for a significant fraction of SN
progenitors.  This binary hypothesis for explaining WR stars and
SNe~Ibc has been around longer (Paczy\'{n}ski 1967) than the idea that
stellar winds of single stars remove the H envelope (Conti 1976).
Several studies of the effects of binary RLOF on massive star
evolution have been conducted (e.g., Podsiadlowski et al.\ 1992;
Wellstein \& Langer 1999; Vanbeveren et al.\ 2007; Eldridge et al.\
2008).  It has been difficult to confirm or refute the idea that
binary RLOF dominates the removal of the H envelopes in massive stars
because of uncertainties in the binary fraction as a function of
initial mass (see Kobulnicky \& Fryer 2007) and the large number of
free parameters in binary models.  Also, until very recently (when
mass-loss rates of hot stars have been revised downward), single-star
evolution seemed to provide a sufficiently plausible alternative.  We
argue here that low mass-loss rates of single stars combined with the
large SN~Ibc fraction now {\it demand} that binary RLOF plays a
dominant role for a large fraction of SNe~Ibc.

Figure~\ref{fig:imfB1} shows a simplified scenario that is radically
different from Figure~\ref{fig:imfS}.  It represents the other extreme
where, instead of assuming that all stars shed their H envelopes via
their own winds in single-star evolution, we adopt the opposite
premise that {\it all} SNe~Ibc have lost their H envelopes via RLOF in
binary systems (following Kobulnicky \& Fryer 2007; Fryer et al.\
1998, 1999; Filippenko 1991; Podsiadlowski et al.\ 1992; Eldridge et
al.\ 2008).  To create Figure~\ref{fig:imfB1}, we simply assumed that
the observed fraction of SNe~Ibc, $\sim$26\%, is identical to the
fraction of massive stars that lose their H envelopes in RLOF, and
that the remaining H-bearing SNe are distributed across the full mass
range.  We of course do not know the binary frequency as a function of
initial mass, so for simplicity, this 26\% is then distributed evenly
across all initial masses of SN progenitors.

This simple ``Binary \#1'' scenario has some advantages over the
standard single-star hypothesis, as well as some drawbacks, a follows.

(1) The initial mass range of 8.5--18 M$_{\odot}$ occupied by the
$\sim$48\% of CCSNe that are SNe~II-P is now in much better agreement
with the inferred mass range of RSG progenitors (Smartt 2009).

(2) The mass range of SNe~II-L is in better agreement with recent
detections of progenitors mentioned above, although perhaps somewhat
too high, and it is unclear how the difference between SNe~II-L and
IIb arises naturally in this scenario.

(3) By redistributing the remaining SNe~II over all initial masses,
this scenario allows for SNe~IIn to be associated with the most
massive stars, consistent with their presumed massive LBV progenitors,
and with the pulsational pair instability in the most extreme cases.
This scenario also has the appealing characteristic that the H-rich
sequence II-P $\rightarrow$ II-L/IIb $\rightarrow$ IIn corresponds to
a sequence of single progenitors with increasing mass-loss rate, and
hence SNe with increasing CSM interaction.

(4) Owing to the fact that SNe~Ib are relatively rare, the mass range
of SNe~Ib (including SNe~Ibc-pec) is surprisingly narrow and low, at
only 8.5--12.4 M$_{\odot}$, if they occupy lower masses than SNe~Ic
within the binary zone in Figure~\ref{fig:imfB1}.  There are currently
no direct detections of SN~Ib progenitors.

\begin{figure*}
\begin{center}
\includegraphics[width=4.6in]{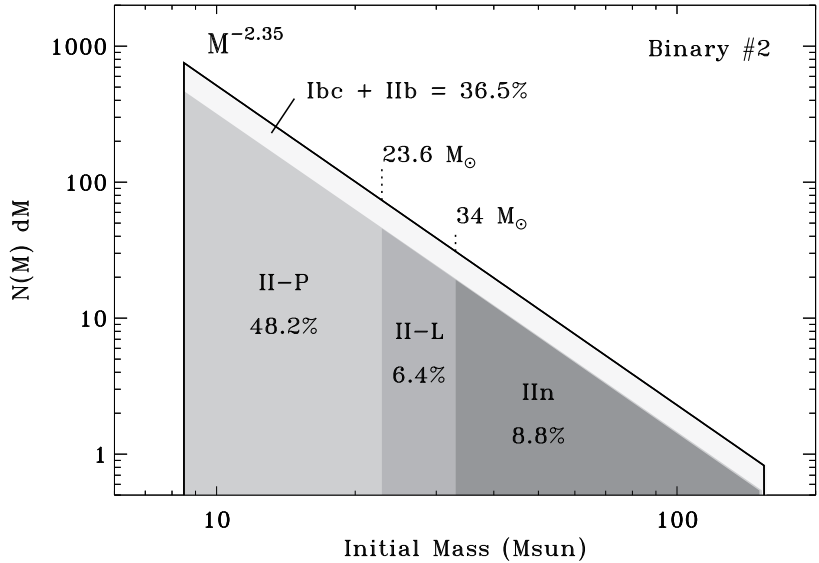}
\end{center}
\caption{Same as Figure~\ref{fig:imfB1}, except that we have included
  SNe~IIb in the same group with SNe~Ibc, all of which are assumed to
  have their envelope stripping dominated by RLOF in close binary
  systems.}
\label{fig:imfB2}
\end{figure*}

A potential objection to this simple binary scenario is that the
fraction of SNe~Ibc compared to SNe~II is observed to be metallicity
dependent (Prantzos \& Boissier 2003; Prieto et al.\ 2008; Boissier
\& Prantzos 2009; Papers I, II, and III), 
as are different WR subtypes (Crowther 2007), and that
SNe~Ic are thought to be associated with massive stars and higher
metallicity because of their specific locations in galaxies (Kelly et
al.\ 2008; Anderson \& James 2009; Papers I and II).  However, this
may still be true even in the simple binary scenario.  Binary RLOF is
only effective at removing the H envelope in most cases, leaving the
He core exposed.  More massive and luminous stars will have stronger
winds with higher radiation-driven mass-loss rates, which will
dominate the {\it subsequent} evolution.  Consequently, only more
massive stars (or perhaps the closest binaries) experience further
significant mass stripping, driving evolution from WN to WC to produce
SNe~Ic rather than SNe~Ib. A corollary is that line-driven winds of WR
stars are metallicity dependent, so while removal of the H envelope
(either by RLOF or LBV-type eruptions) is insensitive to metallicity,
the further evolution from WN to WC (and hence, the production of
SNe~Ic) will be highly dependent on metallicity.

One last complication is that this scenario places SNe~IIb as single
stars and SNe~Ib as binaries.  This leaves us without a satisfying
explanation as to why such a tiny difference in surface H mass
separates SNe~IIb and SNe~Ib (e.g., Chornock et al.\ 2010; Elmhamdi et
al.\ 2006; Filippenko et al.\ 1994), which otherwise look extremely
similar, and it ignores observational results suggesting that the
progenitor of the nearby Type~IIb SN~1993J was most likely a binary
system (Aldering et al.\ 1994; Maund et al.\ 2004; Maund \& Smartt
2009).  There is also evidence for binarity in the case of SN~2001ig
(Ryder et al.\ 2004, 2006; Silverman et al.\ 2009; Maund et al.\
2007), which was also Type~IIb.  These issues motivate the alternative
binary scenario discussed next.

\subsection{Dominated by Close Binaries  \#2}

The second binary-dominated progenitor scenario that we consider is
similar to the first, except that we now include all SNe~IIb along
with SNe~Ib and Ic as stars that lose their H envelopes primarily
through binary RLOF.  The motivation for this, as explained above, is
the close morphological relationship between SNe~IIb and Ib ---
SNe~IIb essentially {\it are} Type~Ib except for a small amount of H
at early times --- plus the observational evidence of the progenitor
of SN~1993J and models for its evolution that are suggestive of a
binary system (Podsiadlowski et al.\ 1993; Aldering et al.\ 1994; 
Maund et al.\ 2004; Maund \& Smartt 2009).

Figure~\ref{fig:imfB2} shows how the IMF could be divided according to
observed SN fractions if we assume that all ``stripped-envelope SNe,''
now including SNe~IIb along with SNe~Ibc, arise from binary RLOF.
Including SNe~IIb as binary systems has three main consequences
compared to the Binary \#1 scenario:

(1)  The fraction of all CCSNe progenitors that lose their H envelopes
through binary RLOF is higher, at $\sim$37\% instead of $\sim$26\%.
Note that both cases Binary \#1 and \#2 imply rather high binary fractions,
as the stripped-envelope progenitors are mainly the mass losers in
RLOF binary systems, but the implied close binary fraction is within
reason (see Kobulnicky \& Fryer 2008).

(2) The upper mass bound for SN~II-P progenitors is shifted to higher
masses (23.6 M$_{\odot}$).  This upper bound is somewhat troublesome,
as it exceeds the 95\% confidence upper limit of 21 M$_{\odot}$
derived from the properties of SN~II-P progenitors (Smartt et al.\
2009).

(3) Most significantly, the mass range for SNe~Ib shifts to higher
mass progenitors than in the Binary \#1 scenario. Assuming that
progenitors of SNe~Ib are less massive than SNe~Ic in the Binary \#1
scenario would dictate that SNe~Ib arise from initial masses of
8.5--12.4 M$_{\odot}$; as noted above, this is low and quite narrow.
If we assume the same for the Binary \#2 hypothesis, but also add the
assumption that SNe~IIb, in turn, are less massive than SNe~Ib, then
the corresponding ranges of initial masses would be 8.5--11
M$_{\odot}$ for SNe~IIb, 11--16 M$_{\odot}$ for SNe~Ib, and $>$16
M$_{\odot}$ for SNe~Ic.  This is an improvement over the Binary \#1
scenario in that it pushes the dividing mass between SNe~Ib and Ic to
higher masses, although 16 M$_{\odot}$ still seems quite low for WR
stars that we expect to shed their own He envelopes via line-driven
winds.  This is remedied in the ``hybrid'' scenario discussed next. An
important caveat is that the monotonic transition SNe~IIb
$\rightarrow$ Ib $\rightarrow$ Ic with increasing initial mass is
probably not strict, as it also depends on initial binary separation
(i.e., very close binaries can remove all of the H and even He layers
in RLOF).  Thus, SNe~IIb could extend to higher masses than 11
M$_{\odot}$ if they arise in relatively wide binaries, for example
(see below).

By dividing SN types into two different and distinct channels
corresponding to single stars and binaries, the Binary \#2 hypothesis
has the appealing quality that it provides a natural continuity in SN
types within each channel, which is lacking otherwise.  With
increasing levels of envelope stripping due to RLOF followed by WR
wind mass loss, the binary channel gives SNe~IIb $\rightarrow$ Ib
$\rightarrow$ Ic.  There may be a continuum of SN progenitors with
different levels of envelope stripping, probably corresponding to
increasing initial metallicity or luminosity, such that a small amount
of residual H separates SNe~IIb and Ib (e.g., Elmhamdi et al.\ 2006;
Chornock et al.\ 2010), whereas a small difference in He mass may
separate SNe~Ib from Ic.

\begin{figure*}
\begin{center}
\includegraphics[width=4.6in]{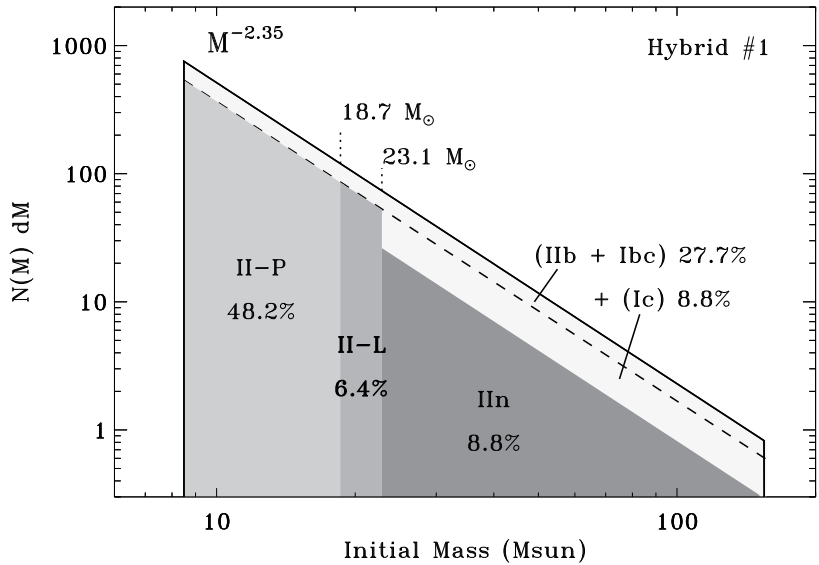}
\end{center}
\caption{Our favoured scenario, combining single and binary star
  evolution.  This is the same as Figure~\ref{fig:imfB2}, except that
  now we have taken roughly half of the SNe~Ic (8.8\% of all CCSNe, to
  match the fraction of SNe~IIn) away from the binary RLOF population
  and mixed them with the single-star population. SNe~Ic that arise
  from single stars are below the dashed line.  Thus, in this scenario
  we assume that half of all single stars above $\sim$23 M$_{\odot}$
  are able to shed their H envelopes via winds or LBV eruptions, while
  the other half retain their H envelopes until just before core
  collapse, producing SNe~IIn.  The difference among the most massive
  stars would depend on the efficiency of winds and LBV eruptions,
  which in turn may depend on properties such as metallicity or
  rotation. The specific numbers shown here are meant to provide just
  one example of a potential hybrid scenario.}
\label{fig:imfH}
\end{figure*}

In the Binary \#2 hypothesis, there is also now a natural continuity
in the single-star channel, giving SNe~II-P $\rightarrow$ II-L
$\rightarrow$ IIn with increasing initial mass and pre-SN mass loss,
and without the puzzling ambiguity between the origins of SNe~IIb and
II-L.  The few direct detections of progenitors that are available
support the notion that the progenitors of SNe~II-L are more massive
than those of SNe~II-P (Elias-Rosa et al.\ 2010b; 2010c), and that
progenitors of SNe~IIn are more massive than SNe~II-L (Gal-Yam \&
Leonard 2009).  The same is true for levels of CSM interaction:
SNe~II-P tend to have extremely weak or undetectable CSM interaction
signatures, SNe~II-L tend to have stronger radio and X-ray emission
(Sramek \& Weiler 1990), and their H$\alpha$ profiles with weak
P-Cygni features are thought to arise from heating of the SN ejecta by
CSM interaction (e.g., Chugai 1991).  SNe~IIn obviously have the
strongest levels of CSM interaction, but there is wide diversity even
among the subclass, with the faintest SNe~IIn like SN~2005ip looking
basically like a SN~II-L with strong narrow emission lines (Smith et
al.\ 2009b), whereas the CSM is opaque and qualitatively changes the
SN in more luminous SNe~IIn such as SN~2006tf and SN~2006gy (Smith et
al.\ 2008b).  The full range for SNe~IIn (34--150 M$_{\odot}$)
encompasses the most luminous RSG that may be responsible for the
fainter SNe~IIn (Smith et al.\ 2009a; see also Yoon \& Cantiello
2010), intermediate cases of SNe~IIn consistent with normal LBVs
(Gal-Yam \& Leonard 2009), as well as the most massive stars with
violent pre-SN mass loss (Smith et al.\ 2007, 2010; Woosley et al.\
2007).

If we relax the requirement that all of the most massive single stars
make successful SNe~IIn, then the lower-right corner of
Figure~\ref{fig:imfB2} provides an attractive parameter space for
massive stars that can collapse to a BH without making a SN display.
If, for example, we allow all {\it single} stars above 50 M$_{\odot}$
in the Binary \#2 scenario to quietly make BHs, then the
redistribution of the remaining mass ranges for SNe~II-P, II-L, and
IIn are still in rough agreement with observational constraints.  Of
course, this would fail to produce the very luminous SNe~IIn that are
thought to come from the most massive stars.

A drawback of this Binary \#2 scenario is that the initial mass range
for SNe~Ic still reaches to uncomfortably low masses, and therefore
dominates most of the mass range for binary progenitors.  Note that if
we allow some of the most massive stars in the binary channel to
undergo quiet BH collapse, we would need to shift the boundary between
SNe~Ic and Ib to even lower initial masses, exacerbating this problem.
Also, the Binary \#2 scenario does not allow any SNe~Ic to come from
single stars, and begs the question of the origin and fate of single
WR stars, which presumably arise from eruptive LBV mass loss in very
massive stars or perhaps through strong winds at super-solar
metallicity.  The next scenario allows some of the most massive single
stars to produce SNe~Ic as well.

\begin{figure*}
\begin{center}
\includegraphics[width=4.6in]{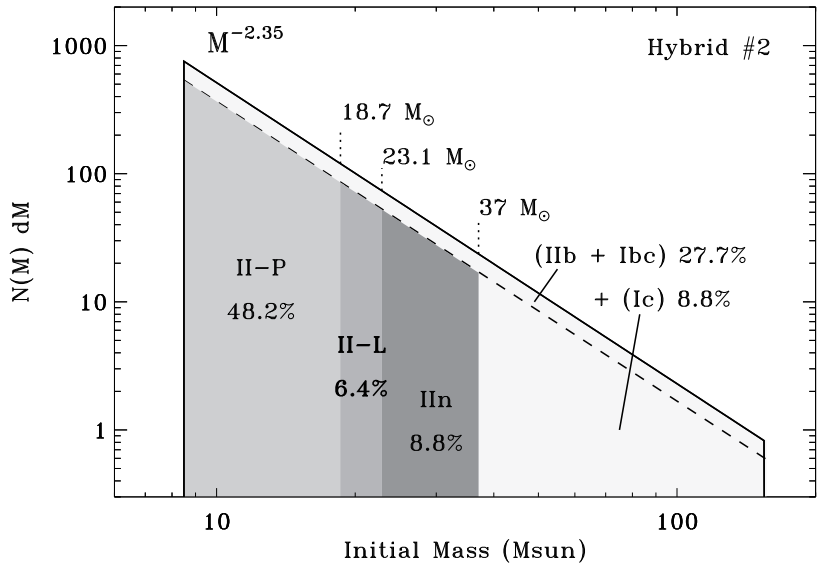}
\end{center}
\caption{This is virtually the same as Figure~\ref{fig:imfH}, except
  that the SNe~IIn and single-star SNe~Ic are not divided equally
  across the range of masses; instead, the SNe~IIn occupy lower
  masses than single-star SNe~Ic.  For equal fractions, the dividing
  mass between SNe~IIn and single-star SNe~Ic would need to be
  $\sim 37$ M$_{\odot}$.}
\label{fig:imfH2}
\end{figure*}

\subsection{A Hybrid Scenario}

One can, of course, play this game {\it ad nauseum} by adjusting the
fraction of SN progenitors that experience binary RLOF, and
redistributing the remainder among single stars in various ways.
Figure~\ref{fig:imfH} shows an example of one ``hybrid'' scenario,
which is a compromise between the standard view of single-star
evolution and the Binary \#2 scenario.  Here we have assumed that
roughly half of the SN~Ic population (we take a fraction equal to
8.8\% of all CCSNe for convenience, equal to the SNe~IIn fraction) may
arise from single-star evolution, while the remainder of SNe~Ic form
via binary RLOF along with SNe~Ib (including Ibc-pec) and SNe~IIb as
before, so that the binary RLOF fraction is 28\% in this hypothetical
scenario.  The binary fraction may be somewhat different or may be
mass dependent, and one can adjust a version of Figure~\ref{fig:imfH}
accordingly to match precise values; the goal here is to be
conceptual.

Although such a scenario may seem more complicated and somewhat {\it
  ad hoc}, it is well motivated, and balances several competing
factors.  Among the most massive stars with initial masses above 23
M$_{\odot}$, it allows single stars to die as {\it either} SNe~Ic or
IIn.  This may be the case if the efficiency of single-star mass loss
depends on additional factors such as rotation or metallicity.  One
can imagine, for example, that very massive stars may be unable to
shed their H envelopes if low metallicity or slower initial rotation
rates weaken their winds or tame the LBV instability.  Under these
circumstances, massive stars might then die as SNe~IIn if they suffer
core collapse while still in the process of attempting to shed their H
envelopes.  Indeed, we noted in Paper~II that SNe~IIn tend to prefer
smaller, lower metallicity galaxies.  The remainder of more rapidly
rotating single stars or higher metallicity single stars might
successfully shed their H envelopes via winds or LBV eruptions and die
as SNe~Ic.  LBV eruptions do seem to be more catastrophic among the
most massive stars (Smith \& Owocki 2006).


Aside from being hypothetical, this scenario has no obvious
disadvantages in view of our knowledge of SN progenitors, and it has
some strengths as follows.

(1)  It maintains very good agreement between the mass range of
SNe~II-P and the inferred mass range of directly detected RSG
progenitors (Smartt 2009).  Putting some of the SNe~Ic back into the
single-star channel has the consequence that it lowers the upper mass
bound required for SNe~II-P compared to the Binary \#2 scenario, improving 
the agreement with observations.  Obviously, we could have chosen the
fraction of SNe~Ic to be a little larger in order to precisely match
the upper mass range for SNe~II-P.

(2) The mass range of SNe~II-L, albeit narrow, is entirely consistent
with known progenitors of this class mentioned earlier.

(3) Figure~\ref{fig:imfH} allows SNe~IIn to arise from among the most
massive stars, consistent with their hypothesised LBV or pulsational
pair instability progenitors.  As in the Binary \#1 and \#2 scenarios,
it provides for the apparent continuity in pre-SN mass loss from
SNe~II-P to II-L to IIn.  The initial mass range of SNe~IIn
progenitors is roughly 23--150 M$_{\odot}$, commensurate with the
known initial mass range of LBVs (Smith et al.\ 2004).  We show an
alternative version of a hybrid scenario in Figure~\ref{fig:imfH2},
wherein we separate SNe~IIn and single-star SNe~Ic by mass, instead of
dividing them half-and-half across all single-star masses above 23
M$_{\odot}$.  This is very similar in principle to the original
standard single-star hypothesis (Figure~\ref{fig:imfS}), but with
SNe~IIb, Ib, and some Ic now excluded as binaries.  In
Figure~\ref{fig:imfH2}, the dividing mass between SNe~IIn and
single-star SNe~Ic is $\sim$36 M$_{\odot}$.  This has the advantage
that classical LBV eruptions above this mass can account for the mass
loss to produce SNe~Ic, but it has the disadvantages that it does not
allow SNe~IIn to arise from the most massive stars, and it does not
allow for other factors like luminous SNe~IIn preferring low
metallicity, or rapid rotation working across a range of masses.  For
these reasons, we tend to favour Figure~\ref{fig:imfH} over
Figure~\ref{fig:imfH2}, but the truth may be somewhere in between.
Differentiating between these two possibilities is difficult, since we
do not yet know how to distinguish single-star from binary SNe~Ic.

(4) As in the Binary \#2 scenario, SNe~IIb arise in binaries,
consistent with the progenitor of SN~1993J (see above).  The initial
mass range of SN~IIb progenitors in this scenario, if they occupy the
low-mass end of RLOF binaries, would be 8.5--12 M$_{\odot}$.  This is
admittedly quite low, and perhaps lower than expected for the
progenitors of SN~1993J ($\sim$15 M$_{\odot}$; Young et al.\ 2006) and
the SN~IIb that gave rise to Cas~A (Krause et al.\ 2008; Rest et al.\
2008), given the strong N enrichment in its CSM (Chevalier \& Kirshner 
1978; Fesen \& Becker 1991; Chevalier \& Oishi 2003).  An
alternative interpretation may be that initial rotation rates,
metallicity, or especially binary separation also play a role here, so
that some of the SN~IIb and SN~Ib progenitors overlap in mass range up to
25 M$_{\odot}$ depending on these conditions.  The wider mass range
would allow more diversity in the progenitors of SNe~IIb, consistent
with the expectations of Chevalier \& Soderberg (2010).  Still,
studies thus far have revealed no surviving companion star for Cas~A
(Thorstensen et al.\ 2001; Krause et al.\ 2008), so there may be
exceptions where some massive single stars produce SNe~IIb as well.
On the other hand, we note that Podsiadlowski et al.\ (1992) expect
cases where the original secondary star that gains mass in RLOF may
experience accelerated evolution and explode first, leaving a widowed
SN~IIb or SN~Ib progenitor to explode as an apparently single
stripped-envelope star.  Perhaps something like this occurred in
Cas~A.

(5) The hybrid scenario gives an appealing explanation for the tiny
observed differences between SNe~IIb and Ib (Elmhamdi et al.\ 2006;
Chornock et al.\ 2010), as in the Binary \#2 scenario.  The initial
masses corresponding to SNe~Ib (and SNe~Ibc-pec) would then be roughly
12--25 M$_{\odot}$.  These are massive stars in binaries whose winds
can get rid of the remaining H, but are not strong enough to fully
remove the He envelope, probably because they are underluminous after
RLOF.  The SNe~Ib progenitors likely correspond to a population of
lower luminosity, early-type WN stars that are difficult to detect
next to their overluminous mass-gainer companions.  Perhaps these
post-RLOF systems would appear as peculiar Be or B[e]-like stars
(mainly due to their overluminous H-rich companions) in nearby
galaxies, likely showing signs of asymmetric CSM.

(6) It retains the quality that SNe~Ic will still trace the most
massive stars, especially those at higher metallicity, whether they
arise from binaries or single stars.  It also gives two different
channels for making SNe~Ic, perhaps providing an avenue for explaining
the diversity among SNe~Ic (i.e., normal vs.\ broad-lined SNe~Ic).
This is an important point beyond the scope of this paper, but
Figure~\ref{fig:imfH} suggests some interesting possibilities.  Even
some broad-lined SNe~Ic, however, appear to arise from only moderately
massive stars, based on the ejecta mass estimates and progenitor
limits (Iwamoto et al.\ 1994; Mazzali et al.\ 2002; Sauer et al.\
2006; Crockett et al.\ 2007).

(7) SNe~Ib, on the other hand, do not trace the highest-mass stars or
regions of high metallicity quite as well in this scenario, since it
is probably the lower mass stars or lower metallicity stars that fail
to drive away their He envelopes.  This scenario would predict
noticeable differences between the environments and progenitors of
SNe~Ib and Ic, with SNe~Ic tending to trace higher initial mass and
higher metallicity.  There is some empirical support for this (Kelly
et al.\ 2008; Anderson \& James 2009; Papers I and II), but further
study should treat SNe~Ib and Ic separately.

(8) SNe~Ib are less common and there are fewer well-studied examples
compared to SNe~Ic, but a recent detailed investigation of the SN~Ib
2007Y revealed a small ejecta mass that suggested a low initial mass
of only 10--13 M$_{\odot}$ for the progenitor, and interestingly,
deduced a progenitor mass-loss rate of only $\la$10$^{-6}$ M$_{\odot}$
(Stritzinger et al.\ 2009).  This mass-loss rate derived from radio
and X-ray data is quite low compared to mass-loss rates of classical
WR stars, supporting the idea that SNe~Ib arise from lower-mass stars
than classical WR stars, and that they have relatively low luminosity
and weak winds (see also Filippenko 1991).  It is even possible, for
instance, that the wind of the mass-gainer companion (e.g., an OB
supergiant) will be stronger than the wind of the SN~Ib progenitor
star, and that the SN blast wave will interact mostly with its
companion's wind.  Whether or not the wind is H-poor is difficult to
ascertain from radio or X-ray observations, and deriving a progenitor
mass-loss rate depends also on an assumed wind velocity (i.e., it may
be significantly lower for a slow B-supergiant wind than for a fast WR
wind).

(9)  There may be a regime where SNe~Ic and SNe~IIn overlap, coming
from the transition between very massive single stars that are
successful in shedding their H envelopes through LBV eruptions (Smith
\& Owocki 2006) and those that cannot.  This may depend on initial
rotation or metallicity, and we speculate that the transition may be
the origin of some of the unusual ``hybrid'' SNe that have been
classified as Type Ia/IIn, such as SNe~2002ic, 2005gj, 1997cy, and
1999E (Turatto et al.\ 2000; Germany et al.\ 2000; Hamuy et al.\ 
2003; Wood-Vasey et al.\ 2003; Rigon et al.\ 2003; Wang et al.\ 2004; 
Chugai \& Yungelson 2004; Kotak et al.\ 2004; Aldering et al.\ 2006; 
Benetti et al.\ 2006; Chugai \& Chevalier 2006; Prieto et al.\ 2007).
Benetti et al.\ (2006) have argued that these may in fact be SNe~Ic
that appear as SNe~IIn because of CSM interaction, rather than SNe~Ia;
this point is speculative and still debated, however.  We conjecture
that unusual SNe~Ibn like SN~2006jc (e.g., Pastorello et al.\ 2007; 
Foley et al.\ 2007) may fit in a similar transitional
category of very massive stars.

All things considered, we favour a hybrid scenario like
Figure~\ref{fig:imfH} as the basic explanation for the observed
fractions of various SN types in large galaxies, invoking binary RLOF
to account for most SNe~IIb, Ib, and some Ic, and yet retaining
single-star mass loss with increasing mass to account for SNe~II-P,
II-L, IIn, and some Ic in the most extreme cases. We stress, however,
that this is hypothetical, with specific binary fractions and other
parameters adopted to encapsulate only the broad properties of various
SN types.  Figure~\ref{fig:imfH} adopted a constant fraction of
progenitors that go through RLOF, whereas this may obviously depend on
initial mass, and RLOF efficiency may depend on other factors like
binary separation and metallicity.  Thus, the mass divisions between
various types are meant as a general guide, rather than definitive
values.  This is certainly an oversimplification, and there may well
be exceptions for individual cases or extreme conditions.  More study
is needed, including detailed population synthesis models with both
binary evolution and LBV-like mass loss for massive stars.  The binary
fraction and its variation with initial mass are key parameters, as is
the behaviour of wind and eruptive mass loss with metallicity and
rotation.  However, we hope that keeping a scenario such as
Figure~\ref{fig:imfH} in mind will be useful to guide intuition for
mapping SNe to stellar initial masses.

\section{Consequences and Future Tests}

If SNe~IIb really result from a different channel than other SNe~II, a
simple comparison of the relative numbers of SNe~Ibc and SNe~II
(including SN~IIb with other SNe~II) is probably misleading.  Such a
comparison would make sense in the standard single-star scenario where
all stars above some threshold mass, $M_{\rm WR}$, make WR stars and
SNe~Ibc (Figure~\ref{fig:imfS}), but we have argued that this simple
hypothesis is contradicted by SN observations.  Instead, an analysis
that retains SN types in-line with the separate binary and single-star
channels discussed here would be more appropriate.  For example,
whereas envelope stripping via binary RLOF should not necessarily
depend on metallicity or initial mass (unless the close binary
fraction changes with mass), the transition SN~IIb $\rightarrow$ Ib
$\rightarrow$ Ic is caused directly by the line-driven wind of the
post-RLOF WR-like star (i.e., proceeding from a low-luminosity WN with
some H, to a normal WN, to WC).  We should therefore be very
interested to see how fractions of these {\it subtypes} change with
metallicity.

Thus, previous studies that have compared the ratio N$_{\rm
  Ibc}$/N$_{\rm II}$, lumping SNe~IIb together with other SNe~II, may
produce somewhat misleading trends and may inspire erroneous
conclusions. In future studies, as larger numbers of all types of
SNe become available, it will be useful to compare relative numbers of
individual subtypes (IIb:Ib:Ic) as well as the ratio of larger groups
that represent different channels [e.g., (IIb + Ib + Ic) / (II-P +
II-L + IIn)] with metallicity and host-galaxy environment.  The
properties of SN~II environments would be particularly interesting;
the Binary \#2 or hybrid hypotheses would predict, for example, that
SNe~IIn come from more massive stars and should therefore trace
clusters and H~{\sc ii} regions to a higher degree than SNe~II-P and
II-L.  This would not be so noticeable for the single-star scenario
shown in Figure~\ref{fig:imfS}.  We would not necessarily expect,
however, that SNe~IIn would be concentrated in galaxy centers, as that
may betray a high-metallicity effect, which leads instead to SNe~Ic
for the most massive stars.  Very massive stars that retain their H
envelopes until shortly before core collapse might instead favour lower
metallicity, and hence smaller host galaxies. This does indeed seem
to be the case, as we point out in Paper~II.

A central hypothesis is that SNe~Ib trace a population of moderately
massive stars that have lost their H envelopes primarily via binary
RLOF.  These progenitors are like classical WR stars in that they are
H deficient, but they differ in that they are likely to be
underluminous with relatively weak winds, and stem from a lower range
of initial masses of roughly 12--25 M$_{\odot}$.  These may not be
recognised as WR stars because of their weaker winds and less
prominent emission lines (see also Filippenko 1991).  In nearby
stellar populations, the SN~Ib progenitors may be among the group of
underluminous early WN stars, or they may reside in binary systems
where they are hard to detect next to their overluminous mass-gainer
companions.  We have speculated that these post-RLOF systems of
moderate mass may appear as Be or B[e]-like stars, perhaps with
asymmetric CSM.  Other potential SN~Ib or Ic progenitor systems are
famous WR+OB systems like V444 Cygni, $\gamma^2$ Vel, or RY~Sct.

Although SNe~Ib are relatively rare, it will be important to
distinguish SNe~Ib from Ic in future analyses, and to clarify their
different properties as well as any range in parameter space where
they may overlap.  It will be especially important to further clarify
the residual He surface mass that separates SNe~Ib from Ic; there may
obviously be examples of a smooth transition in He mass between them.
If SNe~Ib arise from RLOF in binary systems, then the mass-loss rates
of the progenitor stars derived from radio and X-ray observations may
be tricky to interpret.  For example, we noted that the wind of the
overluminous mass-gainer companion may be stronger than the SN
progenitor star itself, and so interaction between the SN blast wave
and the companion's wind might dominate the observed radio and X-ray
emission.  Without a radiative shock to produce strong Balmer lines,
it would be difficult to determine whether the wind is deficient in
hydrogen.\footnote{In the special case of SN~2006jc, the dense CSM
  produced strong He~{\sc i} lines, so one can infer that the
  progenitor star suffered a precursor eruption and that the CSM was
  not from a companion (Pastorello et al.\ 2007; Foley et al.\ 2007;
  Smith et al.\ 2008a); other cases of SNe~Ib are less clear.}

Lastly, it would be interesting to further investigate differences
among environments and progenitors of SNe~Ic, since this class alone
makes up 15\% of all CCSNe (a substantial fraction of the most
massive stars), and may have multiple progenitor channels.  Do the
broad-lined SNe~Ic arise preferentially from one channel?  This is a
key question in regard to the progenitors of long-duration gamma-ray 
bursts.

\section{Conclusions}

We have studied the observed fractions of different SN types from
LOSS, and considered the implications for massive star evolution.
Assuming a Salpeter IMF, we have examined what ranges of initial mass
are needed to account for the observed fractions of SNe~II-P, II-L,
IIn, IIb, Ib, and Ic under various assumptions about the roles of
stellar winds and close binary RLOF in stellar evolution.  We briefly
list the main conclusions here, which apply to stellar evolution in
relatively large galaxies.

(1)  A major finding is that the high observed fraction
of SNe~Ibc cannot be reconciled with predictions of single-star
evolution, where a star's own wind dominates the removal of its H
envelope.  The initial-mass range corresponding to the observed
population of classical WR stars can only account for about half of the
observed SNe~Ibc, so classical WR stars are not the progenitors of a
significant fraction of SNe~Ibc.  Similarly, the initial mass above
which single stars are expected to shed their own envelopes provides a
vastly insufficient fraction of stripped-envelope progenitors, even
with the overly generous mass-loss rates adopted in most stellar
evolution models.

(2)  Instead, we find it likely that RLOF in binary systems is
responsible for the stripped-envelope progenitors of most SNe~IIb and
Ib, and probably a large fraction of SNe~Ic as well.  If these are
distributed over the full range of masses, then SNe~IIb and Ib
probably arise from lower initial masses of 8.5--25 M$_\odot$, and 
SNe~Ic arise from more massive stars with stronger winds.

(3)  Even if binary RLOF dominates the removal of the H envelope, the
further removal of the He layer depends on metallicity-dependent
line-driven winds of the WR star, so SNe~Ic are still expected to
favour more luminous stars and higher metallicity environments.

(4) If the progenitors of SNe~Ib and IIb are not classical WR stars
because their initial masses are too low, then what kind of stars are
the progenitors?  We conjecture that they are probably underluminous
H-poor stars with weak winds that would not necessarily be recognised
as WR stars with prominent emission-line spectra.  They may be easily
hidden by their overluminous mass-gainer companions, which may in some
cases appear as B[e] supergiants or related stars with asymmetric CSM.
If so, one must be cautious when interpreting signatures of the CSM
interaction in SNe~Ib, as the emission may in some cases be dominated
by SN shock interaction with a companion star's wind.

(5) If binary RLOF is important in producing stripped-envelope
progenitors that are a substantial fraction (1/4 to 1/3) of all SN
progenitors, then it would be a mistake to use statistics of SN types
or WR/O-star ratios to guide models for single-star evolution.

(6) After shifting most stripped-envelope progenitors to the binary
RLOF channel, the progenitors of the remaining SN types (H-rich single
stars, wide binaries, and possibly mass-gainers in RLOF binaries) must
be redistributed across the full range of initial masses.  In our
favoured scenario (Figure~\ref{fig:imfH}), SNe~II-P correspond to
initial masses of roughly 8.5--18 M$_{\odot}$, SNe~II-L to 18--23
M$_{\odot}$, and SNe~IIn to 23--150 M$_{\odot}$.  This produces good
agreement with mass ranges inferred from progenitor studies of
SNe~II-P, II-L, and IIn.  In particular, this allows some SNe~IIn to
arise from among the most massive stars, as suggested by some very
luminous SNe~IIn.  Most stellar evolution models fail to account for
very massive stars reaching core collapse without shedding their H
envelope, but this is an expected outcome of the lower mass-loss rates
now dictated by observations.  We also find it likely that some
fraction of the most massive single stars shed their H envelopes to
produce WR stars and SNe~Ic, probably due to high metallicity.  This
allows for the possibility that SNe~IIn favour low-metallicity
environments.

(7) We briefly consider the possibility that some massive stars
collapse directly to BHs without a visible SN display.  We can rule
out this option for the scenario of standard single-star evolution,
because it would make all the problems we note with
Figure~\ref{fig:imfS} worse.  We find no empirical support for the
argument that the ``RSG problem'' may imply direct SN-less BH
formation, because this problem largely goes away with more reliable
SN subtype fractions and with the realization that some RSG stars
evolve to other types of progenitors before exploding.  Though we
cannot rule out the possibility that some massive single stars within
a particular mass range suffer quiet collapse to a BH in a
binary-dominated scenario, quiet BH collapse is not required to
explain the observed relative fractions of CCSNe.

\smallskip\smallskip\smallskip\smallskip
\noindent {\bf ACKNOWLEDGMENTS}
\smallskip
\scriptsize

We are grateful to the many students, postdocs, and other
collaborators who have contributed to the Katzman Automatic Imaging
Telescope and the Lick Observatory Supernova Search over the past two
decades.  We thank the Lick Observatory staff for their assistance
with the operation of KAIT.  A.V.F.'s group has been supported by many
grants from the US National Science Foundation (NSF; most recently
AST-0607485 and AST-0908886), the TABASGO Foundation, US Department of
Energy SciDAC grant DE-FC02-06ER41453, and US Department of Energy
grant DE-FG02-08ER41653. KAIT and its ongoing operation were made
possible by donations from Sun Microsystems, Inc., the Hewlett-Packard
Company, AutoScope Corporation, Lick Observatory, the NSF, the
University of California, the Sylvia \& Jim Katzman Foundation, and
the TABASGO Foundation.

\end{document}